\documentclass[11pt]{article}
\usepackage{amsmath, amsthm, amssymb}
\usepackage{float,epsfig}
\usepackage{graphicx}
\usepackage{rotating}
\usepackage{subfigure}
\usepackage{caption}
\usepackage{subcaption}
\usepackage{tkz-graph}
\usepackage{color}
\usepackage{algorithm}
\usepackage{multicol}
\usepackage[utf8]{inputenc}
\usepackage{braket}
\usepackage{subcaption, multirow}
\usepackage{qcircuit}
\usepackage[overload]{empheq}
\usepackage{hyperref}
\usepackage{xcolor}
\usepackage{color,graphicx,float,lscape,enumerate}
\usepackage{algorithm}
\usepackage[dash,dot]{dashundergaps}
\usepackage{txfonts}

\usepackage[noend]{algpseudocode}
\usepackage{mathtools}

\usepackage{xparse}

\usepackage{lipsum}



\begin{document}

\newtheorem{theorem}{\bf Theorem}[section]
\newtheorem{proposition}[theorem]{\bf Proposition}
\newtheorem{definition}[theorem]{\bf Definition}
\newtheorem{corollary}[theorem]{\bf Corollary}
\newtheorem{example}[theorem]{\bf Example}
\newtheorem{exam}[theorem]{\bf Example}
\newtheorem{remark}[theorem]{\bf Remark}
\newtheorem{lemma}[theorem]{\bf Lemma}
\newcommand{\nrm}[1]{|\!|\!| {#1} |\!|\!|}

\newcommand{\calL}{{\mathcal L}}
\newcommand{\calX}{{\mathcal X}}
\newcommand{\calY}{{\mathcal Y}}
\newcommand{\calZ}{{\mathcal Z}}
\newcommand{\calW}{{\mathcal W}}
\newcommand{\calA}{{\mathcal A}}
\newcommand{\calB}{{\mathcal B}}
\newcommand{\calC}{{\mathcal C}}
\newcommand{\calK}{{\mathcal K}}
\newcommand{\C}{{\mathbb C}}
\newcommand{\Z}{{\mathbb Z}}
\newcommand{\R}{{\mathbb R}}
\renewcommand{\SS}{{\mathbb S}}
\newcommand{\LL}{{\mathbb L}}
\newcommand{\st}{{\star}}
\def\kernel{\mathop{\rm kernel}\nolimits}
\def\sigan{\mathop{\rm span}\nolimits}

\newcommand{\klasse}{{\boldsymbol \Delta}}

\newcommand{\ba}{\begin{array}}
\newcommand{\ea}{\end{array}}
\newcommand{\von}{\vskip 1ex}
\newcommand{\vone}{\vskip 2ex}
\newcommand{\vtwo}{\vskip 4ex}
\newcommand{\dm}[1]{ {\displaystyle{#1} } }

\newcommand{\be}{\begin{equation}}
\newcommand{\ee}{\end{equation}}
\newcommand{\beano}{\begin{eqnarray*}}
\newcommand{\eeano}{\end{eqnarray*}}
\newcommand{\inp}[2]{\langle {#1} ,\,{#2} \rangle}
\def\bmatrix#1{\left[ \begin{matrix} #1 \end{matrix} \right]}
\def\cmatrix#1{\left( \begin{matrix} #1 \end{matrix} \right)}
\def \noin{\noindent}
\newcommand{\evenindex}{\Pi_e}

%\newcommand {\proof} {\par{\it Proof}. \ignorespaces}
%\newcommand {\eproof}
%      {\sigace
%        {\ \vbox{\hrule\hbox{\vrule height1.3ex\hskip0.8ex\vrule}\hrule}}
%        \par}

%%%%%%%%%%%%%%%%%%%%%%%%%%%%%%%%%%%%%%%%%%%%%%%%%%%%%%%%%%%%%%%%%%%%%%%%%%

\def \R{{\mathbb R}}
\def \C{{\mathbb C}}
\def \F{{\mathbb F}}
\def \K{{\mathbb K}}
\def \H{{\mathbb H}}
\def \cu{\mathrm{CU}}

\def \T{{\mathbb T}}
\def \Pb{\mathrm{P}}
\def \N{{\mathbb N}}
\def \Ib{\mathrm{I}}
\def \Ls{{\Lambda}_{m-1}}
\def \Gb{\mathrm{G}}
\def \Hb{\mathrm{H}}
\def \Lam{{\Lambda}}

\def \Qb{\mathrm{Q}}
\def \Rb{\mathrm{R}}
\def \Mb{\mathrm{M}}
\def \norm{\nrm{\cdot}\equiv \nrm{\cdot}}

\def \A{{{\mathbb P}_1(\C^{n\times n})}}
\def \H{{\mathbb H}}
\def \L{{\mathbb L}}
\def \G{{\F_{\tt{H}}}}
\def \S{\mathbb{S}}
\def \s{\mathbb{s}}
\def \sigmin{\sigma_{\min}}
\def \elam{\Lambda_{\epsilon}}
\def \slam{\Lambda^{\S}_{\epsilon}}
\def \Ib{\mathrm{I}}
\def \Tb{\mathrm{T}}
\def \d{{\delta}}

\def \Lb{\mathrm{L}}
\def \N{{\mathbb N}}
\def \Ls{{\Lambda}_{m-1}}
\def \Gb{\mathrm{G}}
\def \Hb{\mathrm{H}}
\def \Delta{\triangle}
\def \Rar{\Rightarrow}
\def \p{{\mathsf{p}(\lam; v)}}

\def \D{{\mathbb D}}

\def \tr{\mathrm{Tr}}
\def \cond{\mathrm{cond}}
\def \lam{\lambda}
\def \sig{\sigma}
\def \sign{\mathrm{sign}}

\def \ep{\epsilon}
\def \diag{\mathrm{diag}}
\def \rev{\mathrm{rev}}
\def \vec{\mathrm{vec}}

\def \ham{\mathsf{Ham}}
\def \herm{\mathsf{Herm}}
\def \sym{\mathsf{sym}}
\def \odd{\mathsf{sym}}
\def \en{\mathrm{even}}
\def \rank{\mathrm{rank}}
\def \pf{{\bf Proof: }}
\def \dist{\mathrm{dist}}
\def \rar{\rightarrow}

\def \rank{\mathrm{rank}}
\def \pf{{\bf Proof: }}
\def \dist{\mathrm{dist}}
\def \Re{\mathsf{Re}}
\def \Im{\mathsf{Im}}
\def \re{\mathsf{re}}
\def \im{\mathsf{im}}

\def \sym{\mathsf{sym}}
\def \sksym{\mathsf{skew\mbox{-}sym}}
\def \odd{\mathrm{odd}}
\def \even{\mathrm{even}}
\def \herm{\mathsf{Herm}}
\def \skherm{\mathsf{skew\mbox{-}Herm}}
\def \str{\mathrm{ Struct}}
\def \cnot{\mathrm{CNOT}}
\def \eproof{$\blacksquare$}

\def \bS{{\bf S}}
\def \cA{{\cal A}}
\def \E{{\mathcal E}}
\def \X{{\mathcal X}}
\def \cH{\mathcal{H}}
\def \cJ{\mathcal{J}}
\def \tr{\mathrm{Tr}}
\def \range{\mathrm{Range}}
\def \adj{\star}
%\newcommand {\proof} {\par{\it Proof}. \ignorespaces}
%\newcommand {\eproof}
    %  {\sigace
        %{\ \vbox{\hrule\hbox{\vrule height1.3ex\hskip0.8ex\vrule}\hrule}}
        %\par}

\def \pal{\mathrm{palindromic}}
\def \palpen{\mathrm{palindromic~~ pencil}}
\def \palpoly{\mathrm{palindromic~~ polynomial}}
\def \odd{\mathrm{odd}}
\def \even{\mathrm{even}}
\def \QT{{\texttt{QT}}}

\newcommand{\algrule}[1][.2pt]{\par\vskip.5\baselineskip\hrule height #1\par\vskip.5\baselineskip}

\newcommand{\tm}[1]{\textcolor{magenta}{ #1}}
\newcommand{\tre}[1]{\textcolor{red}{ #1}}
\newcommand{\tb}[1]{\textcolor{blue}{ #1}}
\newcommand{\tg}[1]{\textcolor{green}{ #1}}
%%%%%%%%%%%%%%%%%%%%%%%%%%%%%%%%%%%%%%%%%%%%%%%%%%%%%%%%%%%%%%%%%%%%%%%%%%%%%%%%

\title{Random sampling of permutations through quantum circuits}
\author{Bibhas Adhikari\\ Fujitsu Research of America, Inc.\\ Santa Clara, California, USA \\
badhikari@fujitsu.com
}

\date{}

\maketitle
\thispagestyle{empty}

% As a general rule, do not put math, special symbols or citations
% in the abstract or keywords.
\begin{abstract}
 In this paper, we introduce a classical algorithm for random sampling of permutations, drawing inspiration from the Steinhaus-Johnson-Trotter algorithm. Our approach takes a comprehensive view of permutation sampling by expressing them as products of adjacent transpositions. Building on this, we develop a quantum analogue of the classical algorithm using a quantum circuit model for random sampling of permutations. As an application, we present a quantum algorithm for the two-sample randomization test to assess the difference of means in classical data. Finally, we propose a nested corona product graph generative model for symmetric groups, which facilitates random sampling of permutations from specific sets of permutations through a quantum circuit model.
\end{abstract}

% Note that keywords are not normally used for peerreview papers.

% For peer review papers, you can put extra information on the cover
% page as needed:
% \ifCLASSOPTIONpeerreview
% \begin{center} \bfseries EDICS Category: 3-BBND \end{center}
% \fi
%
% For peerreview papers, this IEEEtran command inserts a page break and
% creates the second title. It will be ignored for other modes.
%\IEEEpeerreviewmaketitle

\section{Introduction}
% The very first letter is a 2 line initial drop letter followed
% by the rest of the first word in caps.
% 
% form to use if the first word consists of a single letter:
% \IEEEPARstart{A}{demo} file is ....
% 
% form to use if you need the single drop letter followed by
% normal text (unknown if ever used by the IEEE):
% \IEEEPARstart{A}{}demo file is ....
% 
% Some journals put the first two words in caps:
% \IEEEPARstart{T}{his demo} file is ....
% 
% Here we have the typical use of a "T" for an initial drop letter
% and "HIS" in caps to complete the first word.
Random sampling of permutations is one of the fundamental problems in combinatorics and computer science that have found applications in many areas including statistical testing, cryptography, and algorithm design. According to \cite{sedgewick1977permutation}, generating all permutations on a set is ``one of the first nontrivial nonnumeric problems to be attacked by computer". A great many algorithms are proposed in the literature for generating permutations of distinct objects, see \cite{knuth2011art} for a list with historical notes. Some well-known algorithms include: Fisher-Yates (Knuth) Shuffle algorithm, Steinhaus-Johnson-Trotter Algorithm, and the Heap's Algorithm. Besides, there are certain recursive algorithms, and randomized algorithms based on Fisher-Yates shuffle are proposed for specific contexts or additional constraints. The literature lacks significant advancements in the quantum circuit synthesis of arbitrary permutations and the development of circuit models for the random sampling of permutations. These concepts are crucial for constructing numerous algorithms within the domain of post-quantum cryptography, see \cite{budroni2024sok} and the references therein. Moreover, generating permutations on $2^n$ elements using quantum circuits could play a vital role in implementing the quantum permutation pad, which establishes a protocol for universal quantum-safe cryptography \cite{kuang2022quantum}.

%The  Fisher-Yates algorithm works by iterating through a list of $N\geq 2$ elements and swaps each element with another (uniformly) randomly chosen element that comes after it (or itself). This method runs in $O(N)$ time, making it highly efficient \cite{knuth2011art}. The Steinhaus-Johnson-Trotter algorithm, also known as the plain changes algorithm, generates all permutations on a set written as an array with transposing adjacent elements successively \cite{knuth2011art}. While the time complexity for generating all permutations is $O(N!)$, each permutation can be obtained from the previous one in $O(1)$ amortized time.  The Heap’s algorithm is another systematic method for generating permutations which works by recursively swapping elements to generate permutations, ensuring that each permutation is produced exactly once, which can be adapted for random sampling by randomly selecting the next element to swap during the recursion. The time complexity for generating all permutations with this algorithm is $O(N!)$, however like the Steinhaus-Johnson-Trotter algorithm, each permutation can be generated with minimal changes from the previous one that requires additional operations to randomize the order. 

%In the recent developments of extending classical methods into the quantum computing domain, leading to the design of quantum algorithms that leverage the principles of quantum mechanics to several real world applications, 

In the context of quantum computation, when classical data is encoded through probability amplitudes of an $n$-qubit quantum state, the permutations should be performed on $N=2^n$ elements to generate permutationally equivalent quantum states that result in a permutation of the classical data \cite{heredge2024permutation}. 
Consider an $n$-qubit quantum state defined as $\ket{\psi}_N = \sum_{j=0}^{N-1} a_j \ket{j}_N$, where the probability amplitudes $a_j$ satisfy the normalization condition $\sum_{j=0}^{N-1} |a_j|^2 = 1$ and $\{\ket{j}_N : j=0,\hdots, N-1\}$ is the canonical basis of the $n$-qubit Hilbert space. Under the action of a permutation $\pi$ on the index set $\{0, 1, \dots, N-1\}$, the evolved quantum state becomes $\pi\ket{\psi}_N = \sum_{j=0}^{N-1} a_{\pi(j)} \ket{j}_N$. This state is said to be permutationally equivalent to $\ket{\psi}_N$, as it preserves the amplitude distribution but reorders the association between amplitudes and basis states.
%Here, an $n$-qubit quantum state defined by $\ket{\psi}_N=\sum_{j=0}^{N-1} a_j\, \ket{j}_N,$ $\sum_{j=0}^{\tb{N-1}} |a_j|^2=1$,  where the coefficients $a_j$ are the probability amplitudes of the computational basis states when evolved under a permutation $\pi$ on $\{0,1,\hdots, N-1\}$, the resulting quantum state is given by $\pi\ket{\psi}_N=\sum_{j=0}^{N-1} a_{\pi(j)}\, \ket{j}_N,$ which is permutationally equivalent to $\ket{\psi}_N$. 
Thus, permutations on $N$ elements given by $a_j,$ $0\leq j\leq N-1,$ whose corresponding permutation matrices are  unitary matrices of order $N\times N,$ can be utilized for quantum state preparation through a quantum circuit which implements a permutation matrix.

Note that the action of a permutation $\pi$ on the quantum state $\ket{\psi}_N$ can also be expressed as $\pi \ket{\psi}_N = \sum_{j=0}^{N-1} a_j \ket{\pi(j)}_N$. This formulation highlights that a permutation acts by reordering the $2^n$ standard basis states, effectively permuting their positions within the superposition. A special class of permutations is given by  $n(n-1)/2$ SWAP gates on $n$-qubit systems that permute the positions of qubits on the register. It is evident that these SWAP gates generate a group of $n!$ permutations that is isomorphic to the symmetric group on $n$ elements, and is therefore insufficient to implement arbitrary permutations on $N=2^n$ elements \cite{kuang2022quantum2}. We denote $\mathcal{S}_k$ as the symmetric group of order $k$ i.e. the set of all permutations on $k\geq 2$ symbols $0,1,\hdots,k-1$. 

%In particular, when $k=2^n$ we denote it as $\mathcal{S}_{2^n}$ or $\mathcal{S}_N.$

In a related line of research, the design of quantum circuits for generating uniform superpositions of permutations plays a pivotal role in various applications.  One of the earliest works in this direction is by Barenco et al. \cite{barenco1997stabilization}, where the authors construct quantum circuits to generate uniform superpositions over $\mathcal{S}_k$ to generate symmetric subspaces for stable quantum computation via controlled SWAP gates. Their approach employs 
$k(k-1)/2$ ancillary qubits prepared in  superpositions. More recent contributions include the work by Chiew et al. \cite{chiew2019graph}, who utilize $\log_2 k!=O(k\log_2 k)$ qubits, and by Bärtschi et al. \cite{bartschi2020grover}, who propose circuits with $O(k^2)$ qubits to generate such superpositions. Moreover, permutations of selected qubits within a multi-qubit system are used to define permutationally symmetric states \cite{moroder2012permutationally} \cite{toth2009entanglement}, and the corresponding operators are referred to as qubit permutation matrices \cite{fijany1999quantum}.

Attempts are made in the literature for quantum circuit synthesis of specific permutation matrices of order $2^n$. In particular, quantum circuit design from the perspective of \textit{reversible logic synthesis} attracted a lot of interest, such as synthesis of combinatorial reversible circuits, see \cite{saeedi2013synthesis} for a survey. Decomposition of permutations into tensor products has shown to be an important step in deriving fast algorithms and circuits for digital
signal processing \cite{egner1997decomposing}. In \cite{shende2003synthesis}, some fundamental existential results are proved concerning the synthesis of permutations in $\mathcal{S}_{2^n}$ with or without the use of ancillary qubits in the circuits. For instance, it established that every even permutation is $CNT$-constructible, and there are $\frac{1}{2}(2^n-n-1)!$ $T$-constructible permutations in $\mathcal{S}_{2^n}.$ Here, $C$ stands for CNOT, $N$ for NOT (or $X$ gate), and $T$ stands for Toffoli gate. 
In \cite{fijany1999quantum}, efficient quantum circuits for certain permutation matrices are developed which play a pivotal role in the factorization of the unitary operators that arise in the wavelet transforms and quantum Fourier transform. These wavelet transforms, in turn the permutation matrices are likely to be useful for quantum image processing and quantum data
compression.

Among recent advancements, in \cite{soeken2019compiling}, the authors consider the quantum synthesis of permutation matrices which utilizes Young-subgroup based
reversible logic synthesis in existing physical hardware of superconducting transmon qubits. A family of recursive methods for the synthesis of qubit permutations on quantum computers with limited qubit
connectivity are proposed in \cite{chen2022optimizing}. The permutation group that can be obtained from quantum circuits of CNOT gates is explored in \cite{bataille2022quantum}. In \cite{sarkar2024quantum}, quantum circuits for permutations which are expressed as products of some specific adjacent transpositions are also obtained.

%\tb{We mention here that, in another related direction of research, quantum circuit design for uniform superposition of permutations plays a crucial role in several applications. In this direction, the permutations on $N$ elements are considered as permuting $N$ qubits. One of the first papers in this direction is \cite{barenco1997stabilization}, where the authors design quantum circuits for uniform superposition of permutations to generate a symmetric subspace for stable quantum computation with the aid of controlled SWAP gates. Here, elements of $\mathcal{S}_N$ are designed for $N$-qubit system with $N(N-1)/2$ ancillary qubits prepared inuniform superpositions. In \cite{chiew2019graph}, Chiew et al. used $\log_2N!=O(N\log_2N)$ qubits, and in \cite{bartschi2020grover}  Bärtschi et al. used $O(N^2)$ to design quantum circuits to generate uniform superposition of permutations on $N$ elements. Besides, permuting certain qubits in a multi-qubit system is employed as a tool to define permutationlly symmetric states \cite{moroder2012permutationally} \cite{toth2009entanglement}, and the corresponding  permutation matrices are known as qubit permutation matrices \cite{fijany1999quantum}. }

%In contrast to the fact that Steinhaus-Johnson-Trotter algorithm is not typically used directly for random sampling of a single permutation in the literature, 

The contributions of this paper are as follows.

\begin{itemize}
    \item We present a recursive framework grounded in the Steinhaus–Johnson–Trotter algorithm \cite{steinhaus1979one} \cite{johnson1963generation} \cite{trotter1962algorithm} for generating permutations on $N$ elements $0,1,\hdots,N-1$, demonstrated through a binary tree structure that encodes each permutation as a product of adjacent transpositions $s_j = (j, j+1)$, $j = 0, \ldots, N-2$. This hierarchical representation provides a clear combinatorial interpretation of the algorithm and underpins the design of a random sampling algorithm for permutations, with time complexity of $O(N^2).$ 

    \item We develop a quantum circuit model for the random sampling of permutations on $N$ elements, employing $O(N \log_2 N)$ qubits. Besides, the proposed circuit model for implementing any specific permutation uses only $\lceil \log_2 N \rceil$ qubits. The respective gate complexities of these models are $O(N^3 \log_2 N)$ for random sampling and $O(N^2 \log_2 N)$ for specific permutations. The core approach relies on quantum circuit synthesis of adjacent transpositions. We demonstrate that an adjacent transposition $s_j$ can be implemented using a generalized Toffoli gate for even $j$, while for odd $j$, it requires a generalized Toffoli gate along with either two or $2h$ CNOT gates, depending on whether $x$ in the bit representation of $j = (x,1)$ is even or odd, where $x \in \{0,1\}^{n-1}$ and $h$ denotes the Hamming distance between $x$ and $x+1$. 

    \item We propose a quantum algorithm, implementable via a quantum circuit model, for performing the two-sample randomization test for the difference of means \cite{good2013permutation} \cite{ernst2004permutation}. This hypothesis test compares the difference in mean values between two samples drawn from a population of $N$ data points. Classically, generating the necessary sample pairs via permutations comprising two disjoint subsets of sizes $K$ and $N-K$ incurs a time complexity of $O(N \cdot {N \choose K})$. Leveraging the proposed quantum circuit model for random sampling of permutations, the proposed quantum algorithm for the two-sample randomization test in the case where $N = 2^n$ and $K = 2^{n - m}$ achieves a time complexity improvement by a factor of up to $O(\sqrt{N})$ over the classical approach.
    
  \item We introduce a graph generative model based on the corona product, termed the \textit{nested corona product graph}, to provide a structured graph-theoretic representation of symmetric groups. Building upon this construction, we develop a quantum circuit model, extending our framework for random sampling of permutations on $n$-qubit systems, to enable sampling from restricted subsets of permutations corresponding to specific subgraphs or vertex-induced subgraphs within the nested corona product graph.  

\end{itemize}

The rest of the paper is organized as follows. In Section \ref{Sec:classical}, we propose new classical algorithms for random sampling of permutations. The Section \ref{sec:qcAT} includes quantum circuit constructions of adjacent transpositions, which are used in Section \ref{Sec:qcforperm} to define a quantum circuit model for random sampling of permutations for $n$-qubit systems. A quantum algorithm for two-sample randomization test is also included in \ref{Sec:qcforperm}. Finally, in Section \ref{Sec:corona}, we introduce a corona product graph representation of symmetric groups and a quantum circuit model for random sampling from its specific subgraphs.

%\IEEEPARstart{T}{his} demo file is intended to serve as a ``starter file'' for IEEE journal papers produced under \LaTeX\ using IEEEtran.cls version 1.8b and later.
% You must have at least 2 lines in the paragraph with the drop letter
% (should never be an issue)
%I wish you the best of success.

%\hfill mds
 
%\hfill August 26, 2015

\section{Generation of permutations through adjacent transpositions}\label{Sec:classical}

% needed in second column of first page if using \IEEEpubid
%\IEEEpubidadjcol

In this section we devise a method to drive explicit decomposition of a permutation on a set $\{0,1,\hdots,N-1\}, N\geq 2$ as a product of adjacent transpositions. This proposed method is based on the popular  Steinhaus-Johnson-Trotter algorithm, which can be implemented in time $O(1)$ per visited permutation \cite{knuth2011art}, see also \cite{wells1961generation} \cite{trotter1962algorithm} \cite{johnson1963generation} \cite{steinhaus1979one} \cite{cardinal2023combinatorial}, \cite{surhone2010steinhaus}. 

Note that there are two commonly used representations of a permutation $\pi$ on $N$ elements. When written in \textit{square brackets}, known as the permutation array notation, a permutation is denoted as $[x_0, x_1, \ldots, x_{N-1}] = \pi$ to indicate that $\pi(j) = x_j$ for $j \in \{0, \ldots, N-1\}$. On the other hand, when written in \textit{parentheses}, referred to as cycle notation, a permutation is expressed as $(x_{j_1}, x_{j_2}, \ldots, x_{j_k})$, meaning that $\pi(x_{j_l}) = x_{j_{l+1}}$ for $1 \leq l \leq k-1$ and $\pi(x_{j_k}) = x_{j_1}$, where $\{j_1, \ldots, j_k\} \subseteq \{0, 1, \ldots, N-1\}$. We use both notations in this work, with the intended meaning clear from the context.

\subsection{Steinhaus-Johnson-Trotter algorithm}

%\tre{In general, a transposition ordering is concerned with any two consecutive permutations differ in a swap of two entries of the permutation. In particular, Steinhaus-Johnson-Trotter ordering generation of permutations is performed by adjacent transpositions inductively. Assuming the permutations known for $N-1,$ say $\pi_j,$ $1\leq j\leq (N-1)!,$ a permutation for $N$ symbols is constructed by replacing $\pi_j$ by a sequence of permutations obtained by inserting the new symbol $N$ in all positions in $\pi_j$ from right to left for each $j$.}

In general, a \textit{transposition ordering} refers to a sequence of permutations in which each pair of consecutive permutations differs by a transposition of two adjacent elements. Specifically, the Steinhaus–Johnson–Trotter algorithm generates permutation arrays based on this principle, where each successive permutation is obtained by swapping two adjacent entries of the previous one. Given all the permutation arrays of $N-1$ elements, denoted by $\pi_j$ for $1 \leq j \leq (N-1)!$, the permutation arrays of $N$ elements can be constructed inductively. For each $\pi_j$, the new symbol $N$ is inserted into every possible position of $\pi_j$, proceeding from right to left. This systematic insertion ensures that all $N!$ permutations are generated without repetition.

For example, the permutations for $N=2$ are obtained as $[0,\tre{1}]_1$ and $[\tre{1},0]_2$ by placing the new symbol $1$ to the right and left of $0.$ Then 
 for $N = 3$ we obtain $[0,1,\tre{2}]_1$, $[0,\tre{2},1]_1,$ $[\tre{2},0,1]_1,$ $[1,0,\tre{2}]_2,$ $[1,\tre{2},0]_2,$ $[\tre{2},1,0]_2.$ Next, for $N = 4$, we have $[0,1,2,\tre{3}]_1,$ $[0,1,\tre{3},2]_1$, $[0,\tre{3},1,2]_1,$ $[\tre{3},0,1,2]_1,$ $[0,2,1,\tre{3}]_2,$ $[0,2,\tre{3},1]_2,$ $[0,\tre{3},2,1]_2$, $[\tre{3},0,2,1]_2,$ $[2,0,1,\tre{3}]_3,$ $[2,0,\tre{3},1]_3,$ $[2,\tre{3},0,1]_3,$ $[\tre{3},2,0,1]_3,$ $[1,0,2,\tre{3}]_4,$ $[1,0,\tre{3},2]_4,$ $[1,\tre{3},0,2]_4$, $[\tre{3},1,0,2]_4$, $[1,2,0,\tre{3}]_5$, $[1,2,\tre{3},0]_5,$ $[1,\tre{3},2,0]_5,$ $[\tre{3},1,2,0]_5,$ $[2,1,0,\tre{3}]_6,$ $[2,1,\tre{3},0]_6$, $[2,\tre{3},1,0]_6,$ $[\tre{3},2,1,0]_6.$ Here the index $j$ for $[x_1,\hdots,x_N]_j$, $x_k\in\{0,\hdots,N-1\}$ signifies that this permutation is obtained from the $j$-th permutation on $N-1$ symbols in the Steinhaus-Johnson-Trotter ordering. The red colored symbol indicates the movement of the $N$-th symbol from right to left. 
 
% \tre{Also, note that we write this permutation in \tb{\textit{square bracket}} \tb{or an array} since it should not be mistaken as the standard notation of cycle permutation, which we denote using  \tb{\textit{parenthesis}}.  Thus $[x_0,x_1,\hdots,x_{N-1}]=\pi$ if $\pi(j)=x_j,$ $j\in\{0,\hdots,N-1\};$ whereas $(x_{j_1},x_{j_2},\hdots,x_{j_k})$ represents the permutation $\pi$ defined by $x_{j_{l+1}}=\pi(x_{j_l}),$ $1\leq l\leq k-1$ and $\pi(x_{j_k})=x_{j_1},$ $\{j_1,\hdots,j_k\}\subseteq \{0,1,\hdots,N-1\}.$ }

\begin{figure}[htbp]%{$t$\textwidth}
\centering
\begin{tikzpicture}
\draw [fill] (0, 0) circle [radius = .07];
\draw [fill] (1.5, 0) circle [radius = .07];
\draw [fill] (3, 0) circle [radius = .07];
\draw [fill] (3.5, 0) circle [radius = .02];
\draw [fill] (3.7, 0) circle [radius = .02];
\draw [fill] (3.9, 0) circle [radius = .02];
\draw [fill] (4.1, 0) circle [radius = .02];
\draw [fill] (4.5, 0) circle [radius = .07];
\draw [fill] (6, 0) circle [radius = .07];
\draw [fill] (7.5, 0) circle [radius = .07];
\draw (0,0) -- (1.5,0);
\draw (1.5,0) -- (3,0);
\draw (4.5,0) -- (6,0);
\draw (6,0) -- (7.5,0);
\node [below] at (0,0) {$s_0$};
\node [below] at (1.5,0) {$s_1$};
\node [below] at (3,0) {$s_2$};
\node [below] at (4.5,0) {$s_{k-4}$};
\node [below] at (6,0) {$s_{k-3}$};
\node [below] at (7.5,0) {$s_{k-2}$};
\end{tikzpicture}
\caption{Coxeter graph of $\mathcal{S}_k$.}
\label{fig:coxpath}
\end{figure}
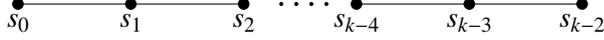

Recall that the symmetric group $\mathcal{S}_{N}$ is a Coxeter group with the generating set of all adjacent traspositions given by $s_j=(j,j+1)$.  The Coxeter graph of $\mathcal{S}_N$ is a path on $N-1$ vertices, each of which represents $s_j,$ $j=0,\hdots,N-2$ from left to right, see Figure \ref{fig:coxpath}. We now discuss an inherent pattern in the Steinhaus–Johnson–Trotter ordering of permutations that enables the explicit decomposition of any permutation as a product of adjacent transpositions $s_j$; see also \cite{johnson1963generation} and the references therein.

For $\mathcal{S}_N,$ $N\geq 2,$  denoting the adjacent transpositions as $s_j=(j,j+1),$ $j=0,1,\hdots,N-1,$ first note that $\mathcal{S}_2=\{I=[0,\tre{1}]_1, [\tre{1},0]_2=s_0\}$. The for $k=3,$ the elements of $\mathcal{S}_3$ are given by
\begin{eqnarray*}
&& [0,1,\tre{2}]_1= I,\,\, [0,\tre{2},1]_1 = s_1, \,\,
 [\tre{2},0,1]_1 = s_1s_0, \\
&& [1,0,\tre{2}]_2 =s_0, \,\, [1,\tre{2},0]_2 =s_0s_1, \,\, [\tre{2},1,0]_2 =s_0s_1s_0.
\end{eqnarray*} In general, it is easy to check that when the elements of $\mathcal{S}_{N-1}$, $N\geq 3$ are expressed as product of adjacent transpositions then the elements of $\mathcal{S}_{N}$ as product of adjacent transpositions can be obtained orderly from Steinhaus-Johnson-Trotter algorithm by positioning the new element $N$ from right to left sequentially at the position $0\leq k\leq N-1$ of the permutation array of $N$ elements. Indeed, the $j$-th element $[x_0,x_1,\hdots,x_{N-2}]_j\in \mathcal{S}_{N-1},$ $1\leq j\leq (N-1)!$, $x_l\in\{0,1,\hdots,N-2\}$, $0\leq l\leq N-2$ drives to obtain $N$ elements of $\mathcal{S}_N$ as $[x_0,x_1\hdots,x_{N-2},\tre{N-1}]_j = [x_0,\, \hdots, x_{N-2}]_j$ and \begin{eqnarray*} && [x_0,\hdots,x_{k-1},\tre{N-1},x_{k},x_{k+1},\hdots,x_{N-2}]_j \\  &&= [x_0,\, \hdots, x_{N-2},\tre{N-1}]_js_{N-2}s_{N-3}\hdots s_{k},\end{eqnarray*} which follows from the fact that when an $s_j=(j,j+1)$ is multiplied from right of an permutation array then the $j$-th and $(j+1)$-th entry of the array are interchanged.

While writing a permutation $\pi$ as product of adjacent transpositions, the length of $\pi$, denoted as $l(\pi)$ is defined as the number of transpositions whose product gives $\pi.$  Then it is natural to consider $l_{\max}(\mathcal{S}_k)=\max\{l(\pi) : \pi\in \mathcal{S}_k\},$ which is well-known to be $k(k-1)/2$ \cite{bjorner2005combinatorics}. In the following lemma, we give a simple proof of this result based on the above discussion. 

\begin{theorem}\label{thm:lengthpi} 
  $l_{\max}(\mathcal{S}_N)=N(N-1)/2$, which is attained by the permutation $[N-1,N-2,\hdots,0].$    
\end{theorem}
\pf %From Theorem \ref{thm:adpermdec}, note that a permutation in $\mathcal{S}_{k-1}$ generates $k$ elements of $\mathcal{S}_k$, $k\geq 3$. 
Note that if $\pi$ is an element of $\mathcal{S}_{k-1}$ then the elements of $\mathcal{S}_k$ that stem from $\pi$ are obtained  by multiplying the adjacent transpositions, represented by the vertices of the Coxeter graph of $\mathcal{S}_k,$ from the right of $\pi$ sequentially. Thus, if $\pi$ denotes an element $\mathcal{S}_{k-1}$ then the length of a permutation $\tau\in \mathcal{S}_k$ obtained from $\pi$ is given by $l(\tau)=l(\pi)+i,$ where $i\in \{0,1,\hdots, k-1\}$ and $\tau\in\{\pi,\pi s_{k-2}, \pi s_{k-2}s_{k-3},\hdots, \pi s_{k-2}s_{k-3}\hdots s_0\}.$ Consequently, the maximum length of an element of $S_k$ is maximum length of an element in $\mathcal{S}_{k-1}$ plus $k-1.$

The length of non-identity permutation in $\mathcal{S}_2$ is $1,$ and hence the maximum length of a permutation in $\mathcal{S}_3$ is $l_{\max}(\mathcal{S}_2)+2=1+2=3.$ Proceeding this way, we obtain $$l_{\max}(\mathcal{S}_k)=1+2+\hdots+(k-1)=\frac{k(k-1)}{2}.$$ 

Thus it follows that the maximum length is obtained by the last permutation obtained through the  Steinhaus-Johnson-Trotter algorithm. \hfill{$\square$}

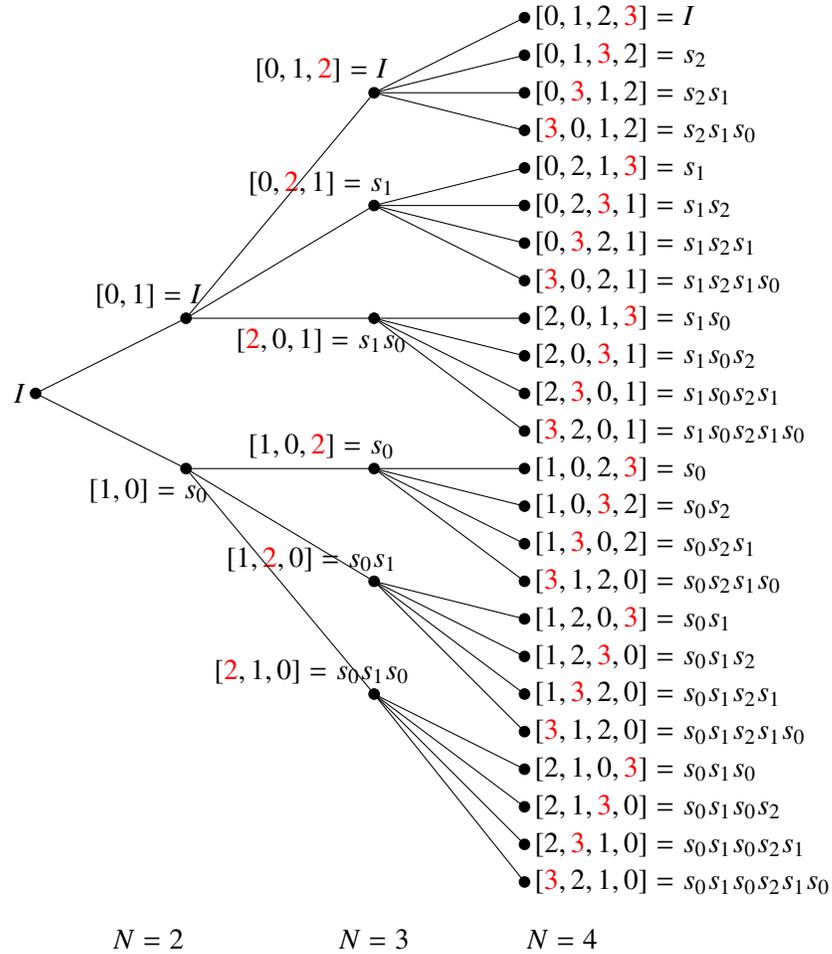
\begin{figure}[htbp]%{$t$\textwidth}
				\centering
				\begin{tikzpicture}
     %\draw (0,0) node[draw,circle] {$t$};
     %\node[draw,circle] at (0,0) {\tb{t}};
%\draw [fill] (-8, 0) circle [radius = .07];
%\draw [fill] (-7, 1) circle [radius = .07];
%\draw [fill] (-7, -1) circle [radius = .07];
%\draw [fill] (-5.5, 2.5) circle [radius = .07];
%\draw [fill] (-5.5, 1.5) circle [radius = .07];
%\draw [fill] (-5.5, 0.5) circle [radius = .07];
%\draw [fill] (-5.5, -0.5) circle [radius = .07];
%\draw [fill] (-5.5, -1.5) circle [radius = .07];
%\draw [fill] (-5.5, -2.5) circle [radius = .07];
     
				\draw [fill] (-0.5, 0) circle [radius = .07];
                \draw [fill] (0.5, 1) circle [radius = .07];
                \draw [fill] (0.5, -1) circle [radius = .07];
                \draw [fill] (2.0, 4.0) circle [radius = .07];
                \draw [fill] (2.0, 2.5) circle [radius = .07];
                \draw [fill] (2.0, 1.0) circle [radius = .07];
                \draw [fill] (2.0, -1.0) circle [radius = .07];
                \draw [fill] (2.0, -2.5) circle [radius = .07];
                \draw [fill] (2.0, -4.0) circle [radius = .07];
                \draw [fill] (3.0, 5) circle [radius = .07];
                 \draw [fill] (3.0, 4.5) circle [radius = .07];
                 \draw [fill] (3.0, 4) circle [radius = .07];
                 \draw [fill] (3.0, 3.5) circle [radius = .07];
                \draw [fill] (3.0, 3) circle [radius = .07];
                \draw [fill] (3.0, 2.5) circle [radius = .07];
                \draw [fill] (3.0, 2) circle [radius = .07];
                \draw [fill] (3.0, 1.5) circle [radius = .07];
                \draw [fill] (3.0, 1) circle [radius = .07];
                \draw [fill] (3.0, 0.5) circle [radius = .07];
                \draw [fill] (3.0, 0) circle [radius = .07];
                \draw [fill] (3.0, -0.5) circle [radius = .07];
                \draw [fill] (3.0, -1) circle [radius = .07];
                \draw [fill] (3.0, -1.5) circle [radius = .07];
                \draw [fill] (3.0, -2) circle [radius = .07];
                \draw [fill] (3.0, -2.5) circle [radius = .07];
                \draw [fill] (3.0, -3) circle [radius = .07];
                \draw [fill] (3.0, -3.5) circle [radius = .07];
                \draw [fill] (3.0, -4) circle [radius = .07];
                \draw [fill] (3.0, -4.5) circle [radius = .07];
                \draw [fill] (3.0, -5) circle [radius = .07];
                \draw [fill] (3.0, -5.5) circle [radius = .07];
                \draw [fill] (3.0, -6) circle [radius = .07];
                \draw [fill] (3.0, -6.5) circle [radius = .07];
				\node [left] at (-0.5, 0) {$I$};
                \node [left] at (0.4, 1.0) {$[0,1]=I$};
				\node [left] at (0.4, -1) {$[1,0]=s_0$};
    \node [left] at (2.0, 4.0) {$[0,1,\tre{2}]=I$};
    \node [left] at (2.0, 2.5) {$[0,\tre{2},1]=s_1$};
				\node [above] at (2.0, 1.0) {$[\tre{2},0,1]=s_1s_0$};
    \node [above] at (1.5, -1.0) {$[1,0,\tre{2}]=s_0$};
				\node [left] at (2.3, -2.5) {$[1,\tre{2},0]=s_0s_1 \,\,\,\,$};
                \node [left] at (2.2, -4.0) {$[\tre{2},1,0]=s_0s_1s_0 \,\,\,\,$};
    \node [right] at (3.0, 5) {$[0,1,2,\tre{3}]=I$};
    \node [right] at (3.0, 4.5) {$[0,1,\tre{3},2]=s_2$};
    \node [right] at (3.0, 4) {$[0,\tre{3},1,2]=s_2s_1$};
    \node [right] at (3.0, 3.5) {$[\tre{3},0,1,2]=s_2s_1s_0$};
    \node [right] at (3.0, 3) {$[0,2,1,\tre{3}]=s_1$};
    \node [right] at (3.0, 2.5) {$[0,2,\tre{3},1]=s_1s_2$};
    \node [right] at (3.0, 2) {$[0,\tre{3},2,1]=s_1s_2s_1$};
    \node [right] at (3.0, 1.5) {$[\tre{3},0,2,1]=s_1s_2s_1s_0$};
    \node [right] at (3.0, 1) {$[2,0,1,\tre{3}]=s_1s_0$};
    \node [right] at (3.0, 0.5) {$[2,0,\tre{3},1]=s_1s_0s_2$};
     \node [right] at (3.0, 0) {$[2,\tre{3},0,1]=s_1s_0s_2s_1$};
    \node [right] at (3.0, -0.5) {$[\tre{3},2,0,1]=s_1s_0s_2s_1s_0$};
    \node [right] at (3.0, -1) {$[1,0,2,\tre{3}]=s_0$};
     \node [right] at (3.0, -1.5) {$[1,0,\tre{3},2]=s_0s_2$};
    \node [right] at (3.0, -2) {$[1,\tre{3},0,2]=s_0s_2s_1$};
    \node [right] at (3.0, -2.5) {$[\tre{3},1,2,0]=s_0s_2s_1s_0$};
    \node [right] at (3.0, -3) {$[1,2,0,\tre{3}]=s_0s_1$};
    \node [right] at (3.0, -3.5) {$[1,2,\tre{3},0]=s_0s_1s_2$};
    \node [right] at (3.0, -4) {$[1,\tre{3},2,0]=s_0s_1s_2s_1$};
    \node [right] at (3.0, -4.5) {$[\tre{3},1,2,0]=s_0s_1s_2s_1s_0$};
    \node [right] at (3.0, -5) {$[2,1,0,\tre{3}]=s_0s_1s_0$};
    \node [right] at (3.0, -5.5) {$[2,1,\tre{3},0]=s_0s_1s_0s_2$};
    \node [right] at (3.0, -6) {$[2,\tre{3},1,0]=s_0s_1s_0s_2s_1$};
    \node [right] at (3.0, -6.5) {$[\tre{3},2,1,0]=s_0s_1s_0s_2s_1s_0$};
    \node [below] at (0.5,-7) {$N=2$};
    \node [below] at (2.0,-7) {$N=3$};
    \node [below] at (3.0,-7) {$N=4$};
        \draw (-0.5,0) -- (0.5,1);
            \draw (-0.5,0) -- (0.5,-1);
            \draw (0.5,1) -- (2.0,4.0);
        \draw (0.5,1) -- (2.0,2.5);
        \draw (0.5,1) -- (2.0,1.0);
        \draw (0.5,-1) -- (2.0,-1.0);
        \draw (0.5,-1) -- (2.0,-2.5);
        \draw (0.5,-1) -- (2.0,-4.0);
        \draw (2.0,4) -- (3.0,5);
        \draw (2.0,4) -- (3.0,4.5);
        \draw (2.0,4) -- (3.0,4);
        \draw (2.0,4) -- (3.0,3.5);
        \draw (2.0,2.5) -- (3.0,3);
        \draw (2.0,2.5) -- (3.0,2.5);
        \draw (2.0,2.5) -- (3.0,2);
        \draw (2.0,2.5) -- (3.0,1.5);
        \draw (2.0,1.0) -- (3.0,1);
        \draw (2.0,1.0) -- (3.0,0.5);
        \draw (2.0,1.0) -- (3.0,0);
        \draw (2.0,1.0) -- (3.0,-0.5);
        \draw (2.0,-1.0) -- (3.0,-1);
        \draw (2.0,-1.0) -- (3.0,-1.5);
        \draw (2.0,-1.0) -- (3.0,-2);
        \draw (2.0,-1.0) -- (3.0,-2.5);
        \draw (2.0,-2.5) -- (3.0,-3);
        \draw (2.0,-2.5) -- (3.0,-3.5);
        \draw (2.0,-2.5) -- (3.0,-4);
        \draw (2.0,-2.5) -- (3.0,-4.5);
        \draw (2.0,-4.0) -- (3.0,-5);
        \draw (2.0,-4.0) -- (3.0,-5.5);
        \draw (2.0,-4.0) -- (3.0,-6);
        \draw (2.0,-4.0) -- (3.0,-6.5);
			%\draw (-2,1) -- (0, 0);
			\end{tikzpicture}
				\caption{The binary tree representation $\mathfrak{P}_N$ of the process of generating all the permutations in terms of product of adjacent transpositions, for $N=2,3,4$.}
				\label{fig:treetransp}
\end{figure}

Next we describe the elements of $\mathcal{S}_N,$ $N\geq 2$ through a binary tree such that the nodes up to order $k\leq N$ represents the permutations in terms of product of the adjacent transpositions $s_j,$ $j=0,1,\hdots,N-2.$ There are $k-1$ nodes of order $k$ stem from a node of order $k-1,$ such that if $\pi$ is the permutations corresponding to the node of order $k-1$ then each node of order $k$ is obtained by multiplying the transpositions $s_{k-2},\hdots,s_0$ sequentially one after one. The binary trees of $\mathcal{S}_N$ for up to $N=4$ is depicted in Figure \ref{fig:treetransp}.

Thus the $\mathcal{S}_N$ can be generated by the adjacent transpositions following a recursive procedure described below. Let \begin{equation}\label{eqn:pik}
    \Pi_k=\left\{I,s_k,s_{k}s_{k-1}, s_ks_{k-1}s_{k-2},\hdots,  s_ks_{k-1}s_{k-2}\hdots s_0\right\},
\end{equation} $0\leq k\leq N-2,$ $|\Pi_k|=k+2.$ The formation of the set $\Pi_k$ can be described using the Coxeter graph $P_{k+1}$ of the symmetric group $\mathcal{S}_{k+2}$ given by  Figure \ref{fig:coxpath}. Indeed, for any $k,$ the set $\Pi_k$ represent the weights of all the directed paths of consecutive lengths from $0$ (representing both the initial and terminal vertex as $s_k$) to $k+1$ with initial vertex $s_k$ to the terminal vertex $s_0$, along with the identity permutation. The weight of such a path is defined as product of all the weights (the adjacent transpositions associated with the vertices) of all the vertices in the path.\\

\noindent{\bf Recursive procedure:} Generation of elements of the symmetric group $S_k,$ $k\geq 2$ \begin{eqnarray}
    && \mathcal{S}_2=\{I,s_0\}=\Pi_0 \nonumber \\
    && \mbox{For} \,\, k> 2, \,\, \mathcal{S}_{k}=\bigcup_{\pi\in \mathcal{S}_{k-1}} \pi \Pi_{k-2},  \label{gensn}
\end{eqnarray}  where  $\pi\, \Pi_{k-2}=\{\pi s: s\in \Pi_{k-2}\}.$

This recursive procedure for generation of all the elements of $\mathcal{S}_N$ using the generating set $S$ can further be demonstrated by a binary tree representation, denoted by $\mathfrak{P}_N$. The root of the tree is considered as the node of order $0$ that denotes the identity permutation on a set with only one symbol. Then there would be $(k+2)$ nodes stem from a node of order $k\in\{0,1,\hdots, N-1\}.$ For $k=1$ i.e. first order nodes represents the identity permutation and $s_0=(0,1).$ In other words, the first-order nodes represent the elements of $\Pi_0.$ For $k\geq 2,$ the $k$-th order nodes of $\mathfrak{P}_N$ represent the elements of $\{\pi \Pi_{k-1} : \pi\in \mathcal{S}_k\}=\mathcal{S}_{k+1}.$ The tree representation of $\mathcal{S}_N, 2\leq N\leq 4$ is exhibited in Figure \ref{fig:treetransp}. Algorithm \ref{algoPerp} provides algorithmic procedure for generation of all the elements of $\mathcal{S}_N.$ Obviously, Algorithm~\ref{algoPerp} can be viewed as an alternative representation of the Steinhaus–Johnson–Trotter algorithm. Consequently, its time complexity for generating all permutations of $N$ elements as permutation arrays is $O(N!)$.  

\begin{algorithm}
\caption{Generation of $\mathcal{S}_N,$ $N\geq 3$  }\label{algoPerp}
\textbf{Input:} $\Pi_k,$ $0\leq k\leq N-2$, $\mathcal{S}_2=\Pi_0$\\
%where $\sigma_j\in \{I,X,Y,Z\},j\in \{1,\hdots,n\}$.\\
%$b=\sum_{j=1}^{n} 2^{n-j}b_j, b\geq 1$ such that $b_j\in\{0,1\}$ and $\sigma_b\in\mathcal{S}_{I,X}^{(n)}$ such that $\sigma_b=\otimes_{j=1}^{n} \sigma_{j}=\sigma_1\otimes\hdots\otimes\sigma_n$, where $\sigma_j=X$ if $b_j=1$ and $\sigma_j=I_2$ if $b_j=0$.  \\
\textbf{Output:} Elements of of $\mathcal{S}_N$ as product of adjacent transpositions 
\begin{algorithmic}
\For{$k=3$ to $N$} 
%\For{$\pi$ in $\mathcal{S}_{k-1}$}
\State $\mathcal{S}_k=\{\pi\Pi_{k-2} : \pi\in \mathcal{S}_{k-1} \}$ 
%\State $\pi=\pi_{0}\pi_1\hdots \pi_{N-2}$
\EndFor
%\EndFor
\State \textbf{End For}
%\State \textbf{End For}
\State \textbf{Return} $\mathcal{S}_N$
\end{algorithmic}
\end{algorithm}

Finally, in Algorithm~\ref{algo14}, we present how to decompose any given permutation array into a product of adjacent transpositions by reversing the steps of the Steinhaus–Johnson–Trotter algorithm. Let us denote $\Pi_k(j)$ as the $j$-th element of $\Pi_k$ (see equation (\ref{eqn:pik})) where $j=1,2,\hdots,k+2.$ Let $\pi=[x_0,x_1,\hdots,x_{N-1}]\in \mathcal{S}_N$ be a given permutation array. The correctness of Algorithm \ref{algo14} can be validated by reversing the construction process of the final output generated by the Steinhaus–Johnson–Trotter algorithm, utilizing the structured organization of $\mathfrak{P}_N$ as a reference framework. As established in the preceding discussion, the position of each element $\alpha\in\{0,1,\hdots, N-1\}$ within $\pi$ uniquely determines the selection of an element from the set $\Pi_{\alpha-1}$ during the stepwise construction of $\mathfrak{P}_N.$ To initiate the verification, consider $\alpha=N-1.$ The position $l_\alpha$ of the element $N-1$ within $\pi$ identifies a specific element in $\Pi_{\alpha-1}.$ Given the entries of $\pi$ are indexed from $0$ to $N-1,$ and the corresponding elements in $\Pi_{\alpha-1}$ are arranged in reverse order, the correspondence is defined by the mapping $l_{\alpha}\mapsto (\alpha+1)-l_{\alpha},$ where $\alpha+1=\left|\Pi_{\alpha-1}\right|.$ Subsequently, this process is repreated iteratively for $\alpha=N-2,N-3,\hdots, 1$ with the array $\pi$ being updated at each step by removing the element $\alpha$ to reflect the reduced configuration. Through this iterative procedure, the selection sequence aligning with the structure of $\mathfrak{P}_N$ is systematically reconstructed, thereby establishing the correctness of Algorithm \ref{algo14}.

%{reversing the construction of a final outcome in the Steinhaus-Johnson-Trotter algorithm and with the aid of the structured formation of $\mathfrak{P}_N$, the correctness of Algorithm \ref{algo14} can be verified as follows. From the discussions above, it is needless to mention that the position of $\alpha \in\{1,\hdots,N-1\}$ in the permutation array $\pi$ decides the choice of an element from $\Pi_{\alpha-1}$ in the construction of $\mathfrak{P}_N.$ First, consider $\alpha=N-1.$ Then the position of $N-1$ as $l_\alpha$  in $\pi$ decides a corresponding element of $\Pi_{\alpha-1}$. However, since the entries of $\pi$ are indexed from $0$ to $N-1,$ and the corresponding elements, which are reversely ordered in $\Pi_{\alpha-1}$, are mapped as $l_{\alpha}\mapsto (\alpha+1)-l_{\alpha},$ where $\alpha+1=\left|\Pi_{\alpha-1}\right|.$ Then for $N-2$ and sequentially decrementing the value of $\alpha,$ the above procedure can be followed by updating $\pi$ at each step with a removal of $\alpha$ from $\pi$ for the next step.}

Since the input permutation $\pi$ is unsorted, locating and removing element $\alpha$ among $\alpha+1$ entries incurs a worst-case cost of $O(\alpha+1)$ per step. Summing over $\alpha\in\{N-1,\hdots,1\}$, the total worst-case complexity of Algorithm \ref{algo14} is $O(N^2)$.

\begin{algorithm}
\caption{Decomposition of a permutation as a product of adjacent transpositions}\label{algo14}
\textbf{Input:} $\pi=[x_0,x_1,\hdots,x_{N-1}]$, $x_i=\pi(i)\in\{0,\hdots,N-1\},$ $0\leq i\leq N-1$; $\Pi_k,$ $k=1,\hdots,N-2$ \\
\textbf{Output:} A decomposition of $\pi$ as a product of adjacent transformations $s_j=(j,j+1)$, $j=0,\hdots,N-2$
\begin{algorithmic}
\For{$\alpha=N-1$ to $1$}
\State{find $l_\alpha$ such that $x_{l_\alpha}=\pi(l_\alpha)=\alpha$ and update $\pi$ by removing $x_{l_\alpha}$ from $\pi$}
%\State{\tb{remove $x_{l_k}$ from $\pi$}}
%\State{\textbf{End For}}
\EndFor
\State{\textbf{End For}}
\State{\textbf{Return} $\prod_{\alpha=1}^{N-1} \Pi_{\alpha-1}\left((\alpha+1)-l_\alpha\right)$}
\end{algorithmic}
\end{algorithm}

For instance, consider $N=4$ and $\pi=[3,2,0,1].$ For $\alpha=N-1=3,$  $l_\alpha=0,$ and $\pi$ update to $[2,0,1].$ For  $\alpha=2,$ $l_{\alpha}=0$ and $\pi$ becomes $[0,1].$ For $\alpha=1,$ $l_{\alpha}=1$ yielding $\pi=[0].$ The output of Algorithm \ref{algo14} is thus $\pi=\Pi_0(2-1)\Pi_1(3-0)\Pi_2(4-0)=\Pi_0(1)\Pi_1(3)\Pi_2(4)=I(s_1s_0)(s_2s_1s_0)=s_1s_0s_2s_1s_0,$ in agreement with the expression given in Figure \ref{fig:treetransp}.

\subsection{Classical algorithm for random sampling of permutations}

%\tre{Now using the tree representation $\mathfrak{P}_N$, we propose an algorithm for sampling a random permutation on $N$ symbols.} 
In this section, we leverage the structure of the tree $\mathfrak{P}_N$, derived from the Steinhaus-Johnson-Trotter algorithm, to formulate an algorithm for sampling permutations from the symmetric group. Note that there are $(k+2)$ nodes that stem from a $k$-th order node of $\mathfrak{P}_N,$ $0\leq k\leq N-1.$ Now we assign a probability $1/(k+2)$ for choosing a node that are originated from of a $k$-th order node. Since there are exactly $N(N -1)(N -2)\cdots 2=N!$ such sequences, each of the distinct sequence produces
a different permutation. Moreover, the probability of choosing a sequence is $\frac{1}{2}\cdot \frac{1}{3}\cdots \frac{1}{N}=1/N!.$ Obviously, the first-order nodes of $\mathfrak{P}_N$ represent the nodes $I$ and $s_0$, which form the set $\{\pi \Pi_0 : \pi\in \mathcal{S}_1\}=\mathcal{S}_2,$ where $S_1$ is the trivial symmetric group on the set containing one symbol only. Similarly, for each node of order $k,$ which represents an element $\pi\in \mathcal{S}_{k-1}$ there are $(k+1)$ nodes stem from each such $\pi$ and these nodes represent the elements of $\pi\Pi_{k-1},$ where for $k\geq 2.$ Thus choosing an element, say $\pi_k$ from $\Pi_k$ uniformly i.e. with probability $1/(k+2)$ for $0\leq k\leq N-2$, we obtain a random permutation $\pi=\pi_0\pi_1\cdots \pi_{N-2}.$ Consequently, we have the Algorithm \ref{algo1}, where 
 $\pi\leftarrow U(A)$ denotes that
the element $\pi$ is sampled uniformly at random from the set $A$. Obviously, there is a scope of simultaneously choosing $\pi_k$ from $\Pi_k$ uniformly in parallel for all $k.$

Now note that the elements of $\Pi_k$ is given in terms of product of adjacent transpositions, and due to Theorem \ref{thm:lengthpi}, in the worst case scenario there would be $N(N-1)/2$ product of adjacent transpositions in the outcome $\pi$ of Algorithm \ref{algo1}. Essentially, the worst-case complexity of random sampling of permutations of $N$ elements due to Algorithm \ref{algo1} is $O(N^2)$ since the time complexity of multiplying $k$ transpositions in a sequence is $O(k)$, where each one requires 
$O(1)$ time.

\begin{algorithm}
\caption{Random sampling from $\mathcal{S}_N,$ $N\geq 3$ through product of adjacent permutations }\label{algo1}
\textbf{Input:} $\Pi_k,$ $0\leq k\leq N-2$, \\
%where $\sigma_j\in \{I,X,Y,Z\},j\in \{1,\hdots,n\}$.\\
%$b=\sum_{j=1}^{n} 2^{n-j}b_j, b\geq 1$ such that $b_j\in\{0,1\}$ and $\sigma_b\in\mathcal{S}_{I,X}^{(n)}$ such that $\sigma_b=\otimes_{j=1}^{n} \sigma_{j}=\sigma_1\otimes\hdots\otimes\sigma_n$, where $\sigma_j=X$ if $b_j=1$ and $\sigma_j=I_2$ if $b_j=0$.  \\
\textbf{Output:} A permutation $\pi\in \mathcal{S}_N$
\begin{algorithmic}
\For{$k=0$ to $N-2$} 
\State{$\pi_k\leftarrow U(\Pi_k)$} 
\State{$\pi=\pi_{0}\pi_1\hdots \pi_{N-2}$}
%\State{\textbf{End For}}
\EndFor
\State{\textbf{End For}}
\State{\textbf{Return} $\pi$}
\end{algorithmic}
\end{algorithm}

Now recall that any $0\leq k\leq N!-1$ can be written as $$k=\sum_{j=0}^{N-1} a_j \, j!, \,\, a_j\in\{0,1,2,\hdots,j-1\}.$$ Since there are $N!$ permutations on a set of $N$ symbols, there is a natural way to associate an integer to a permutation, called a \textit{ranking function} for the permutations \cite{de1967note} \cite{myrvold2001ranking}. Then starting from $0$-th order node of $\mathfrak{P}_N$ the  $j$-th order nodes can be represented by the the ordered numbers $0,1, \hdots, j-1$ that can be chosen with probability $1/j$ that corresponds to elements of the ordered set $\Pi_{j-1},$ $j\geq 1.$ Therefore, choosing a random permutation essentially boils down to choosing a number $a_j$ from the set $\{0,1,\hdots, j-1\}$ that can be done simultaneously for $1\leq j\leq N-1.$ Thus we have an alternative version of the Algorithm \ref{algo1} as described in Algorithm \ref{algo12}. Note that the for loops in both the algorithms are not necessarily required and this step can be parallelized.

%consecutively one after one, we obtain a sequence of numbers $j_1,j_2,\hdots j_{N-1}$, where $1\leq j_l\leq l+1,$ $l\in\{1,\hdots,N-1\}.$

\begin{algorithm}
\caption{(A variant of Algorithm \ref{algo1}))}\label{algo12}
\textbf{Input:} $K_j=\{0,1,\hdots,j-1\}, 1\leq j\leq N-1$, \\
%where $\sigma_j\in \{I,X,Y,Z\},j\in \{1,\hdots,n\}$.\\
%$b=\sum_{j=1}^{n} 2^{n-j}b_j, b\geq 1$ such that $b_j\in\{0,1\}$ and $\sigma_b\in\mathcal{S}_{I,X}^{(n)}$ such that $\sigma_b=\otimes_{j=1}^{n} \sigma_{j}=\sigma_1\otimes\hdots\otimes\sigma_n$, where $\sigma_j=X$ if $b_j=1$ and $\sigma_j=I_2$ if $b_j=0$.  \\
\textbf{Output:} A permutation $\pi\in \mathcal{S}_N$
\begin{algorithmic}
\For{$j=1$ to $N-1$} 
\State{$k_j\leftarrow U(K_j)$} 
%\State{$\pi=\Pi_{0}[j]\pi_1[]\hdots \pi_{N-2}$}
%\State{\textbf{End For}}
\EndFor
\State{\textbf{End For}}
\State{\textbf{Return} $\prod_{j=1}^{N-1} \Pi_{j-1}(k_j+1)$}
\end{algorithmic}
\end{algorithm}

Note that the steps within the \textit{for loop} in Algorithms \ref{algo1} and \ref{algo12} can be implemented in a parallel setup, allowing them to be executed in constant time complexity, $O(1)$. Consequently, the overall time complexity of the algorithms remains $O(N^2)$, as justified by Theorem~\ref{thm:lengthpi}, which asserts that any permutation can be expressed as a product of at most $N(N-1)/2$ adjacent transpositions. A list of existing algorithms for generating random permutations, along with their time complexities, can be found in Chapter 6 of \cite{devroye1986}. The Fisher–Yates algorithm, which operates in linear time $O(N)$, remains the most efficient for classical applications. On the other hand, we demonstrate in later sections that representing permutations as products of adjacent transpositions is particularly well-suited for quantum implementation. 

Table~\ref{Tab:cca} summarizes the asymptotic time complexities of the classical algorithms presented in this section.

\begin{table}[htp]
    \centering
    \begin{tabular}{|c|c|}
         \hline
        Algorithm & Time complexity \\\hline\hline
        Enumeration of all permutations, Algorithm \ref{algoPerp}  & $O(N!)$ \\ \hline
       Decomposing a permutation array as product of  & $O(N^2)$ \\
        adjacent transpositions, Algorithm \ref{algo14} & \\ \hline
       Random sampling of permutations,  & $O(N^2)$\\
       Algorithms \ref{algo1} $\&$ \ref{algo12} & \\
        \hline
    \end{tabular}
    \caption{Asymptotic time complexity of classical algorithms for generating permutation arrays, assuming parallel execution of the \textit{for loops} in Algorithms~\ref{algo1} and~\ref{algo12}}
    \label{Tab:cca}
\end{table}

\section{Quantum circuit implementation of adjacent transpositions for $n$-qubit systems }\label{sec:qcAT}

%\subsection{Preliminaries}\label{Sec:prelim}

\begin{figure}[htbp]
				\centering
				\begin{tikzpicture}
     %\draw (0,0) node[draw,circle] {$t$};
     %\node[draw,circle] at (0,0) {\tb{t}};
				\draw [fill] (0, 0) circle [radius = .05];
				\node [below] at (0, 0) {$\,\,t$};
				\draw [fill] (-2, 1) circle [radius = .1];
			  \node [left] at (-2, 1) {$\ket{\tb{0}}_1$};
              % \node [right] at (-2, 1) {$\,\, \tre{0}$};
				\node [above] at (-1, 0.5) {$0$};
			\draw (-2,1) -- (0, 0);
				\draw [fill] (2, 1) circle [radius = .1];
				\node [right] at (2, 1) {$\ket{\tb{1}}_1$};
   % \node [left] at (2, 1) {$\tre{1}\,\,$};
				\node [above] at (1, 0.5) {$1$};
					\draw (2,1) -- (0, 0);
				\draw [fill] (-3, 2) circle [radius = .1];
				\node [left] at (-3, 2) {$\ket{\tb{0}}_2$};
    %\node [right] at (-3, 2) {$\tre{0}$};
				\node [above] at (-2.5, 1.5) {$0$};
					\draw (-3,2) -- (-2, 1);
			\draw [fill] (-1, 2) circle [radius = .1];
			\node [right] at (-1, 2) {$\ket{\tb{1}}_2$};
   %\node [left] at (-1, 2) {$\tre{1}$};
			\node [above] at (-1.5, 1.5) {$1$};
					\draw (-1,2) -- (-2, 1);
			\draw [fill] (3, 2) circle [radius = .1];
			\node [right] at (3, 2) {$\ket{\tb{3}}_2$};
   %\node [left] at (3, 2) {$\tre{3}$};
			\node [above] at (1.5, 1.5) {$0$};
					\draw (3,2) -- (2, 1);
			\draw [fill] (1, 2) circle [radius = .1];
			\node [left] at (1, 2) {$\ket{\tb{2}}_2$};
			\node [above] at (2.5, 1.5) {$1$};
					\draw (1,2) -- (2, 1);
			\draw [fill] (-3.5, 3) circle [radius = .1];
			\node [above] at (-3.5, 3) {$\ket{\tb{0}}_3$};
			\node [above] at (-3.5, 2.3) {$0$};
					\draw (-3.5,3) -- (-3, 2);
			\draw [fill] (-2.5, 3) circle [radius = .1];
			\node [above] at (-2.5, 3) {$\ket{\tb{1}}_3$};
			\node [above] at (-2.5, 2.3) {$1$};
					\draw (-2.5,3) -- (-3, 2);
			\draw [fill] (-1.5, 3) circle [radius = .1];
			\node [above] at (-1.5, 3) {$\ket{\tb{2}}_3$};
			\node [above] at (-1.5, 2.3) {$0$};
					\draw (-1.5,3) -- (-1, 2);
			\draw [fill] (-0.5, 3) circle [radius = .1];
			\node [above] at (-0.5, 3) {$\ket{\tb{3}}_3$};
			\node [above] at (-0.5, 2.3) {$1$};
					\draw (-0.5,3) -- (-1, 2);
			\draw [fill] (0.50, 3) circle [radius = .1];
			\node [above] at (0.50, 3) {$\ket{\tb{4}}_3$};
			\node [above] at (0.5, 2.3) {$0$};
					\draw (0.5,3) -- (1, 2);
			\draw [fill] (1.5, 3) circle [radius = .1];
			\node [above] at (1.5, 3) {$\ket{\tb{5}}_3$};
			\node [above] at (1.5, 2.3) {$1$};
					\draw (1.5,3) -- (1, 2);
			\draw [fill] (2.5, 3) circle [radius = .1];
			\node [above] at (2.5, 3) {$\ket{\tb{6}}_3$};
			\node [above] at (2.5, 2.3) {$0$};
					\draw (2.5,3) -- (3, 2);
			\draw [fill] (3.5, 3) circle [radius = .1];
			\node [above] at (3.5, 3) {$\ket{\tb{7}}_3$};
			\node [above] at (3.5, 2.3) {$1$};
					\draw (3.5,3) -- (3, 2);
			\end{tikzpicture}
				\caption{The quantum binary tree $\texttt{QT}_3.$ The $i$-th order nodes are labelled by $\ket{\tb{q}}_i=\ket{q_{i-1}\hdots q_1q_0}_i$, where $0\leq q\leq 2^{i}-1,$ $1\leq i\leq 3$ and $q=\sum_{j=0}^{i-1} q_j2^{j}, q_j\in\{0,1\}.$ }
				\label{fig:tree}
\end{figure}
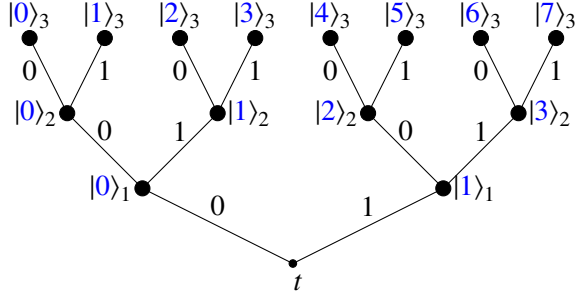

First we recall the Quantum binary Tree, henceforth $\QT_n,$ proposed in \cite{adhikari2024local} to visualize the canonical basis elements of an $i\leq n$-qubit subsystem in an $n$-qubit system through a combinatorial procedure. The terminal nodes (from left to right) of $\QT_n$ represent the ordered canonical basis elements $\ket{q_{n-1}\hdots q_0}_n,$ $q_j\in\{0,1\}$ corresponding to the $q$-th basis element of an $n$-qubit system, where $q=\sum_{j=0}^{n-1} q_j2^{j}$ provides the binary representation $q=(q_{n-1},\hdots,q_1,q_0)$ of $q\in\{0,1,\hdots,N-1\},$ $N=2^n.$ The $\ket{q_j}$ in $\ket{q_{n-1}\hdots q_0}_n$ represents the state of the $(n-j)$-th qubit, $0\leq j\leq n-1.$  The nodes of order $i$ in $\QT_n$ represent the canonical ordered basis elements of an $i$-qubit system, $1\leq i\leq n$. For example, $\QT_3$ is given in Figure \ref{fig:tree}.

In what follows, we derive quantum circuit implementation of adjacent transpositions $s_{j}=(j,j+1)$, $j=0,1,\hdots, 2^n-2$ on $2^n$ elements by utilizing the construcyion of $\QT_n.$ The meaning of $s_j$ in quantum context is the following. Let $\ket{\psi}_n=\sum_{j=0}^{2^n-1} a_j\, \ket{j}_n$ be an $n$-qubit quanum state. Then $$s_j\ket{\psi}_n=\sum_{l=0}^{j-1} a_l\, \ket{l}_n + a_{j+1}\ket{j}_n + a_j\ket{j+1}_n + \sum_{l=j+2}^{2^n-1} a_j\, \ket{l}_n.$$ Thus, applying an adjacent transposition $s_j$ on an $n$-qubit quantum state written as a linear combination of canonical quantum states interchanges the probability amplitudes corresponding to $\ket{j}_n$ and $\ket{j+1}_{n}.$ From the $\QT_n$ perspective, the application of $s_j$ interchanges the probability amplitudes corresponding to a pair of consecutive terminal nodes and the others remain unchanged. 

\subsection{Quantum circuit for the transposition $s_j$ when $j$ is even}\label{sec:cToffoli}

In this section, we show that when $j$ is even, $s_j$ is a generalized Toffoli gate. First we recall the definition of generalized Tofolli gates. Consider \begin{equation}\label{eqn:basis} \left\{\ket{q}_n=\ket{q_{n-1}\hdots q_1q_0}_n : q_j\in\{0,1\}\right\}\end{equation} as the canonical ordered basis of $\C^{2^n}.$ 
%where $k=0,1,\hdots 2^n-1, k=\sum_{j=0}^{n-1} k_j2^{j}.$ 

Then given $x\in\{0,1,\hdots,2^{n-1}-1\}$ with $(n-1)$-bit binary representation $x=(x_{n-2}\hdots x_1,x_0),$ the Toffoli gate $T^{(n)}_x: \C^{2^n} \rightarrow \C^{2^n}$ is defined as
\begin{equation}\label{def:Tgate}
    T^{(n)}_x \ket{q_{n-1}\hdots q_1q_0}_n =\begin{cases}
        \ket{q_{n-1}\hdots q_{1}}_{n-1} (X\ket{q_0}_1) \,\, \mbox {if}\\
         \hfill{(x_{n-2}\hdots x_0)=(q_{n-1}\hdots q_{1})}; \\
        \ket{q_{n-1}\hdots q_0}_n, \,\, \mbox{otherwise.}
    \end{cases}
\end{equation} 
We also use the alternative notation $T^{(n)}_{(x_{n-2},\cdots,x_0)}$ for $T^{(n)}_x.$ If $x=\boldsymbol{1}_{n-1},$ the all-one $(n-1)$-bit string, $T^{(n)}_{\boldsymbol{1}_{n-1}}$ represents the standard $n$-qubit Toffoli gate. Otherwise, it is called generalized Tofolli gate. Then for any $x\in \{0,1,\hdots,2^{n-1}-1\},$ the quantum state $\ket{x}_{n-1}=\ket{x_{n-2}\hdots x_0}_{n-1}$ represents a vertex of order $n-1$ in $\QT_n.$  The quantum states $\ket{x}_{n-1}\ket{0}_1=\ket{x0}_n$ and $\ket{x}_{n-1}\ket{1}_1=\ket{x1}_n$ are the terminal vertices of $\QT_n$ that stem from $\ket{x}_{n-1}.$ Further, for such $x$, $T^{(n)}_x(\ket{x0}_n)=\ket{x1}_n$ and $T^{(n)}_x(\ket{x1}_n)=\ket{x0}_n$, and the other canonical basis elements of $\C^{N}$ remain invariant under $T^{(n)}_x$. Then we have the following theorem.

\begin{theorem}
 Let $x\in\{0,1,\hdots,2^{n-1}-1\}.$ Then $T^{(n)}_x$ represents the adjacent transposition $(2x,2x+1).$   
\end{theorem}
\pf For $x\in\{0,1,\hdots, 2^{n-1}-1\}$ with the binary representation $(x_{n-2},\hdots,x_1,x_0)$ such that $x=\sum_{j=0}^{n-2} x_j2^{j},$ the indices corresponding to the basis states $\ket{x0}_n$ and $\ket{x1}_n$ are given by 
\begin{eqnarray*} && x_{n-2}2^{n-1} + x_{n-3}2^{n-2}+\hdots+x_02^1+0(2^0) = \sum_{j=0}^{n-2} x_{j}2^{j+1} \\
&& = 2\left(\sum_{j=0}^{n-2}x_{j}2^{j}\right)=2x\end{eqnarray*} and 
\begin{eqnarray*} && x_{n-2}2^{n-1} + x_{n-3}2^{n-2}+\hdots+x_02^1+1(2^0) = \sum_{j=0}^{n-2} x_{j}2^{j+1} +1 \\
&& = 2\left(\sum_{j=0}^{n-2} x_{j}2^{j}\right) +1=2x+1,\end{eqnarray*} respectively. Consequently, for any $x\in\{0,1,\hdots,2^{n-1}-1\}$ $T^{(n)}_x (\ket{2x}_n)=\ket{2x+1}_n$ and $T^{(n)}_x (\ket{2x+1}_n)=\ket{2x}_n$, and other basis states in $\C^{N}$  remain invariant under $T^{(n)}_x$. This completes the proof. \hfill{$\square$}

%\tm{\begin{eqnarray*} && x_{n-2}2^n + x_{n-3}2^{n-1}+\hdots+x_02^1+02^0 = \sum_{j=1}^{n-1} x_{j-1}2^j \\&& = 2\left(\sum_{j=1}^n x_{j-1}2^{j-1}\right) = 2\left(\sum_{j=0}^{n-2}x_{j}2^{j}\right)=2x\end{eqnarray*} and  \begin{eqnarray*} && x_{n-2}2^n + x_{n-3}2^{n-1}+\hdots+x_02^1+12^0 = \sum_{j=1}^{n-1} x_{j-1}2^j +1 \\&& = 2\left(\sum_{j=1}^n x_{j-1}2^{j-1}\right) +1=2x+1,\end{eqnarray*}} respectively. Consequently, for any $x\in\{0,1,\hdots,2^{n-1}-1\}$ $T^{(n)}_x (\ket{2x}_n)=\ket{2x+1}_n$ and $T^{(n)}_x (\ket{2x+1}_n)=\ket{2x}_n$, and other basis states in $\C^{N}$  remain invariant under $T^{(n)}_x$. This completes the proof. \hfill{$\square$} 

Thus the generalized Toffoli gates directly implements $s_j,$ when $j$ is even. Now we derive quantum circuit implementation of generalized Toffoli gates with $X$ gate and the standard Toffoli gate.

%\subsubsection{Quantum circuits for generalized Toffoli gates}

Let $S_{I_2,X}^{(n)}=\{\otimes_{j=n-1}^{0} \sigma_j: \sigma_j\in\{I_2,X\}\}$ denote set of $n$-qubit quantum gates that are tensor product of $I_2$ and the Pauli matrix $X.$ Then for $b\in\{0,1,\hdots,2^n-1\}$ with $b=(b_{n-1},b_{n-2},\hdots,b_{0})\in\{0,1\}^{n}$ as the binary string representation of $b,$ we can associate an element $\sigma_b=\otimes_{j=n-1}^0 \sigma_j\in S_{I_2,X}^{(n)}$, where $$\sigma_j=\begin{cases}
    X \,\, \mbox{if} \,\, b_j=1 \\
    I_2 \,\, \mbox{if} \,\, b_j=0
\end{cases}.$$ Consequently, the map $\{0,\hdots,2^n-1\}\rightarrow S^{(n)}_{I_2,X}$ defined as $b\mapsto \sigma_b$ is bijective. Now in what follows, we show that $T^{(n)}_x$ for any $x\in\{0,1,\hdots,2^{n-1}-1\}$ can be constructed by using $T^{(n)}_{\boldsymbol{1}_{n-1}}$ and elements of $S_{I_2,X}^{(n)}$ in the following theorem.  

\begin{theorem}\label{Thm:gTgate}
Let $b\in\{0,1,\hdots,2^n-1\}$ be an integer with $n$-bit representation $(b_{n-1},\hdots,b_1,b_0).$ The $\sigma_bT^{(n)}_{\boldsymbol{1}_{n-1}}\sigma_b=T^{(n)}_{\sum_{j=1}^{n-1}\overline{b_j}2^j}=T^{(n)}_{\overline{b_{n-1}}\hdots \overline{b_2} \, \overline{b_1}},$ which corresponds to the adjacent transposition $(2x,2x+1)$ for the set $\{0,1,\hdots,2^n-1\}$ where $x=\sum_{j=1}^{n-1}\overline{b_j}2^j.$ 
\end{theorem}
\pf First we show that for any $\sigma_b,$ $\sigma_bT^{(n)}_{\boldsymbol{1}_{n-1}}\sigma_b\in \{T_x^{(n)} : 0\leq x\leq 2^{n-1}-1\}.$ Observe that $\sigma_bT^{(n)}_{\boldsymbol{1}_{n-1}}\sigma_b$ non-trivially acts on a basis element $\ket{q}_n=\ket{q_{n-1}\hdots q_1}_{n-1}\ket{q_0}_1$ of $\C^{2^n}$ given by equation (\ref{eqn:basis}), if and only if $q_j=0$ when $b_j=1$ and $q_j=1$ when $b_j=0$, $1\leq j\leq n-1,$ and
\begin{eqnarray*} && \sigma_bT^{(n)}_{\boldsymbol{1}_{n-1}}\sigma_b\ket{q_{n-1}\hdots q_1q_0}_n \\ &=& \sigma_bT^{(n)}_{\boldsymbol{1}_{n-1}}\ket{\boldsymbol{1}_{n-1}}_{n-1}(\sigma_0\ket{q_0}_1) \\
&=&  (\sigma_{n-1}\ket{1}_1) \cdots (\sigma_{1}\ket{1}_1)  (\sigma_0 X\sigma_0\ket{q_0}_1)\\
&=& \ket{q_{n-1}\hdots q_1}_{n-1} (\sigma_0 X\sigma_0\ket{q_0}_1)\\
&=&\ket{q_{n-1}\hdots q_1}_{n-1} \ket{q_0\oplus 1}_1.\end{eqnarray*} This yields
$\sigma_bT^{(n)}_{\boldsymbol{1}_{n-1}}\sigma_b\ket{q}_n=\ket{q+1}_n$ if $q_0=0,$ and $\sigma_bT^{(n)}_{\boldsymbol{1}_{n-1}}\sigma_b\ket{q}_n=\ket{q-1}_n$ if $q_0=1$ for any $q\in\{0,1,\hdots,2^{n}-1\}.$ Therefore, $\sigma_bT^{(n)}\sigma_b$ represents an adjacent transposition for any given $b\in\{0,1,\hdots, 2^n-1\}$. 

%$$\sigma_bT^{(n)}_{\boldsymbol{1}_{n-1}}\sigma_b\ket{k_{n-1}\hdots k_1k_0}_n = \begin{cases}\sigma_bT^{(n)}_{\boldsymbol{1}_{n-1}}\ket{\boldsymbol{1}_{n-1}}\ket{k_0}=\ket{k_{n-1}\hdots k_1}_n\ket{k_0\oplus 1}_1, \,\, \mbox{if} \,\, b_0=0\\\sigma_bT^{(n)}_{\boldsymbol{1}_{n-1}}\ket{\boldsymbol{1}_{n-1}}\ket{k_0\oplus 1}=\ket{k_{n-1}\hdots k_1k_0}_n, \,\, \mbox{if} \,\, b_0=1.\end{cases}$$ 

Moreover, $\sigma_bT^{(n)}_{\boldsymbol{1}_{n-1}}\sigma_b$ transforms the basis element $\ket{\sum_{\substack{j: b_j=0\\ j=1}}^{n-1}2^j + q_0}_n$ into $\ket{\sum_{\substack{j: b_j=0\\ j=1}}^{n-1}2^j + (q_0\oplus 1)}_n,$ $q_0\in\{0,1\}$ and vice-versa, whereas the remaining basis elements remain invariant. Let $\emptyset\neq J=\{j_1,\hdots,j_l\}\subseteq \{1,\hdots,n-1\}$ such that $b_j=0$ if $j\in J.$ Then the adjacent transposition 
$\left(\sum_{j\in J}2^j, \sum_{j\in J}2^j+1\right)$ is described by the Tofolli gate $T^{(n)}_{\sum_{j=1}^{n-1}\overline{b_j}2^j},$ where $\overline{b_j}=b_j\oplus 1,$ $\oplus$ denotes the modulo-$2$ addition.  If $J=\emptyset$ i.e. $b_j=1$ for $1\leq j\leq n-1$ then set $\sum_{j\in J}2^j=0.$ This concludes the proof. \hfill{$\square$}

\begin{algorithm}
  \caption{Circuit construction of generalized Toffoli gates for $n$-qubit systems}
  \begin{algorithmic}[1]
    \State Let $x\in\{0,1,\hdots,2^{n-1}-1\}$ and we want to construct the circuit for $T^{(n)}_x$ (see equation (\ref{def:Tgate})).
    \State Find the $(n-1)$-bit representation $(x_{n-2}\cdots x_0)$ of $x.$
    \State Define an $n$-bit representation $(b_{n-1},b_{n-2},\cdots,b_0)$ for the integer $b$ such that $b_j=x_{j-1}\oplus 1,$ $1\leq j\leq n-1$, and $b_0\in\{0,1\}.$
    \State Define the quantum gate $\sigma_b=\sigma_{n-1}\otimes \sigma_{n-2}\otimes \cdots \otimes \sigma_1\otimes \sigma_0$ such that $\sigma_j=X$ if $b_j=1$ and $\sigma_j=I_2$ if $b_j=0.$
    \State Then $\sigma_bT^{(n)}_{\boldsymbol{1}_{n-1}}\sigma_b$ implements the generalized gate Toffoli gate $T^{(n)}_x.$
  \end{algorithmic}
  \label{Alg:gTgate}
\end{algorithm}

\begin{figure}[htbp]
    \centering
     \subfigure [\centering ]{{\includegraphics[width=0.25\textwidth]{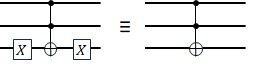} }}%
     %\caption{}
    \qquad
     \subfigure [\centering ]{{\includegraphics[width=0.42\textwidth]{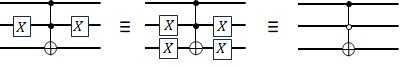} }}%
    \qquad
    \subfigure [\centering ]{{\includegraphics[width=0.42\textwidth]{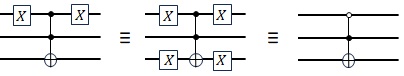} }}%
    \qquad
    \subfigure [\centering ]{{\includegraphics[width=0.42\textwidth]{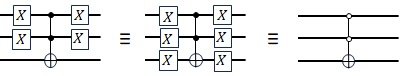} }}%
    % \qquad
   % \subfloat [\centering ]{{\includegraphics[width=0.35\textwidth]{lognormal-dist.png} }}%
\centering \caption {Generalized Toffoli gates for $3$-qubit system through standard Toffoli gate and the $X$-gate. (a) $T_{11}=(6,7)$, (b) $T_{10}=(4,5)$, (c) $T_{01}=(2,3),$ (d) $T_{00}=(0,1)$}
    \label{fig:Togates}
\end{figure}

We provide the Algorithm \ref{Alg:gTgate} which describes the circuit construction of a generalized Toffoli gate using $X$ or NOT gate and the standard Toffoli gate using Theorem \ref{Thm:gTgate}. It should be noted that there are two choices of $\sigma_b$ for a given $x$ with either $b_0=0$ or $b_0=1.$ However, for both the choices it gives the same generalized Tofolli gate. The following remark emphasizes on the number of generalized Toffoli gates. In particular, we demonstrate the construction of generalized Toffoli gates for a $3$-qubit system which is further illustrated by explicit circuit representations in Figure \ref{fig:Togates}.

\begin{remark}
Note that corresponding to two $n$-bit strings, $(b_{n-1},\hdots,b_1,0):=\beta_0$ and $(b_{n-1},\hdots,b_1,1):=\beta_1$ there are two strings of $X$-gates, say $\sigma_{\beta_0}$ and $\sigma_{\beta_1},$ respectively. However, $\sigma_{\beta_0}T^{(n)}_{\boldsymbol{1}_{n-1}}\sigma_{\beta_0}=\sigma_{\beta_1}T^{(n)}_{\boldsymbol{1}_{n-1}}\sigma_{\beta_1}$, that is,   $(\otimes_{n-1}^{1}\sigma_j\otimes I_2)T^{(n)}_{\boldsymbol{1}_{n-1}}(\otimes_{n-1}^{1}\sigma_j\otimes I_2) = (\otimes_{n-1}^{1}\sigma_j\otimes X)T^{(n)}_{\boldsymbol{1}_{n-1}}(\otimes_{n-1}^{1}\sigma_j\otimes X).$ Hence the total number of $\sigma_b$'s is $2^{n-1}$ which is the total number of (generalized) Toffoli gates on $n$-qubits corresponding to $(b_{n-1},\hdots,b_1),$ $b_j\in\{0,1\},$ $1\leq j\leq n-1$ obtained through the above derivation. \end{remark}

For example, consider $n=3.$ Then for $b\in\{0,1,\hdots,7\}$ there are eight binary strings. Then the $2^{3-1}=4$ Toffoli gates on $3$-qubit system can be obtained as follows.
\begin{eqnarray*}
&& \begin{cases}
     b=(\underline{00}0) \\
     b=(\underline{00}1)
 \end{cases}   \mapsto \begin{cases}
     \sigma_b=I_2I_2I_2\\
     \sigma_b=I_2I_2X
 \end{cases} \mapsto T_{2^1+2^0}=T_{3:=x} \\ && \hfill{=T_{(1,1)}=(2x,2x+1)=(6,7);} \\
&&  \begin{cases}
     b=(\underline{01}0) \\
     b=(\underline{01}1)
 \end{cases}   \mapsto \begin{cases}
     \sigma_b=I_2XI_2\\
     \sigma_b=I_2XX
 \end{cases} \mapsto T_{2^1}=T_{2:=x} \\ && \hfill{=T_{(1,0)}=(2x,2x+1)=(4,5);} \\
&& \begin{cases}
     b=(\underline{10}0) \\
     b=(\underline{10}1)
 \end{cases}   \mapsto \begin{cases}
     \sigma_b=XI_2I_2\\
     \sigma_b=XI_2X
 \end{cases} \mapsto T_{2^0}=T_{1:=x} \\ && \hfill{=T_{(0,1)}=(2x,2x+1)=(2,3);} \\
 && \begin{cases}
     b=(\underline{11}0) \\
     b=(\underline{11}1)
 \end{cases}   \mapsto \begin{cases}
     \sigma_b=XXI_2\\
     \sigma_b=XXX
 \end{cases} \mapsto T_{0:=x} \\ && \hfill{=T_{(0,0)}=(2x,2x+1)=(0,1).}
\end{eqnarray*}

%In Figure \ref{fig:Togates} we exhibit the construction of all Toffoli gates for $3$-qubit system through standard Toffoli gate and the $X$-gate.

Thus we prove that, each transposition of the form $(j,j+1),$ when $i\in\{0,1,\hdots,2^n-1\}$ is even, can be obtained using the standard Toffoli gate and a string of one-qubit $X$-gates $\sigma_b.$ Next, we devise quantum circuit implementation of the transpositions $(j,j+1)$  when $i$ is odd as follows. 

\subsection{Quantum circuit for the transposition $s_j$ when $j$ is odd}

%Before proceeding further, first recall that menaing of a transposition as a quantum gate for an $n$-qubit system. Let $U_{(i,i+1)}$ denote the unitary gate (permutation matrix) corresponding to the transposition $(i,i+1).$ Then, for any quantum state $\psi=\sum_{j=0}^{2^n-1}a_j\ket{j}$ expressed as linear sum of canonical basis states as an input for $U_{(i,i+1)},$ we have $$U_{(i,i+1)}\ket{\psi}=\sum_{j=0}^{i-1} a_j\ket{j} + a_{i+1}\ket{i} + a_i\ket{i+1}+\sum_{j=i+2}^{2^n-1} a_j\ket{j}$$ i.e. it represents the transformation $$[a_0,\hdots,a_{i-1},a_i,a_{i+1},a_{i+2},\hdots,a_{2^n-1}]^T \mapsto [a_0,\hdots,a_{i-1},a_{i+1},a_i,a_{i+2},\hdots,a_{2^n-1}]^T.$$  

\begin{figure}[htbp]%{$t$\textwidth}
				\centering
				\begin{tikzpicture}
     %\draw (0,0) node[draw,circle] {$t$};
     %\node[draw,circle] at (0,0) {\tb{t}};
				\draw [fill] (-0.5, 0) circle [radius = .1];
                \draw [fill] (1.5, 0) circle [radius = .1];
                \draw [fill] (3.5, 0) circle [radius = .1];
                \draw [fill] (5.5, 0) circle [radius = .1];
				\node [below] at (-0.5, 0) {$\ket{00}_2$};
                \node [below] at (1.5, 0) {$\ket{01}_2$};
                \node [below] at (3.5, 0) {$\ket{10}_2$};
                \node [below] at (5.5, 0) {$\ket{11}_2$};
				\draw [fill] (-1, 1) circle [radius = .1];
                \draw [fill] (0, 1) circle [radius = .1];
                \draw [fill] (1, 1) circle [radius = .1];
                \draw [fill] (2, 1) circle [radius = .1];
                \draw [fill] (3, 1) circle [radius = .1];
                \draw [fill] (4, 1) circle [radius = .1];
                \draw [fill] (5, 1) circle [radius = .1];
                \draw [fill] (6, 1) circle [radius = .1];
				\node [above] at (-1, 1) {$\substack{\tb{a_0} \\\ket{000}_3}$};
                \node [above] at (0, 1) {$\substack{\tb{a_1} \\ \ket{00\tre{1}}_3}$};
                \node [above] at (1, 1) {$\substack{\tb{a_2} \\ \ket{010}_3}$};
                \node [above] at (2, 1) {$\substack{\tb{a_3} \\ \ket{01\tre{1}}_3}$};
                \node [above] at (3.0, 1) {$\substack{\tb{a_4} \\ \ket{100}_3}$};
                \node [above] at (4, 1) {$\substack{\tb{a_5} \\ \ket{10\tre{1}}_3}$};
                \node [above] at (5, 1) {$\substack{\tb{a_6} \\ \ket{110}_3}$};
                \node [above] at (6, 1) {$\substack{\tb{a_7} \\ \ket{11\tre{1}}_3}$};
            \draw (-0.5,0) -- (-1,1);
            \draw (-0.5,0) -- (0,1);
            \draw (1.5,0) -- (1,1);
            \draw (1.5,0) -- (2,1);
            \draw (3.5,0) -- (3,1);
            \draw (3.5,0) -- (4,1);
            \draw (5.5,0) -- (5,1);
            \draw (5.5,0) -- (6,1);
			%\draw (-2,1) -- (0, 0);
			\end{tikzpicture}
				\caption{The $n$-th and $(n-1)$-th order basis states and their combinatorial connectivity through the quantum binary tree $\texttt{QT}_3.$ The labeling in \tb{blue} for the $n$-th order nodes represents the coefficient for any quantum state $\ket{\psi}=\sum_{j=0}^{2^n-1} \tb{a_j} \ket{j}$ with respect to the canonical ordering of the basis states in $\C^{2\otimes n}.$ The $n$-th qubit corresponding to odd indexed basis states of $\C^{2\otimes n}$ are colored \tre{red}.}
				\label{fig:tree2}
\end{figure}
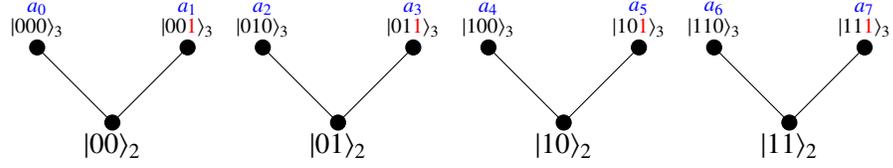

First note that, for any given odd integer $j\in\{0,1,\hdots,2^n-1\},$ the corresponding basis state is given by $\ket{j}_n=\ket{x1}_n,$ which represents a terminal node of $\QT_n$ that stems from the $(n-1)$-th order node in $\QT_n$ representing the quantum state $\ket{x}_{n-1}$ (see Figure \ref{fig:tree2} for $n=3$). Then for $j\in\{0,1,\hdots,2^n-3\},$ consider the basis state $\ket{j+2}_n=\ket{(x+1)1}_n$ which stems from the $(n-1)$-th order node of $\QT_n,$ represented by $\ket{x+1}_{n-1}.$ If $(x_{n-2},\hdots,x_0)$ is the $(n-1)$-bit representation of $x$ then $\ket{x}_{n-1}=\ket{x_{n-2}\hdots x_1x_0}_{n-1}$. If $x$ is an even then $x_0=0$, and  $x_0=1$ if $x$ is odd. Consequently, if $x=(x_{n-2},\hdots, x_1,0)$ is even then $x+1$ is odd with $(n-1)$-bit representation $(x_{n-2},\hdots, x_1,1)$ and hence $\ket{j}_n=\ket{x1}_n=\ket{x_{n-2}\hdots x_101}_n$ and $\ket{j+2}_n=\ket{(x+1)1}_n=\ket{x_{n-2}\hdots x_111}_n.$ Further, if $x=(x_{n-2}\hdots x_11)$ is odd then there exists $l\in\{1,\hdots, n-1\}$ number of bit-places where the binary representations of $x$ and $x+1$ differ, and $l$ is called the Hamming distance of $x$ and $x+1$. For example, if $x=(011)=3$ then $x+1=(100)=4$, and hence the Hamming distance between $x$ and $x+1$ is $3$ in a $3$ bit-system. Now, since $j=(x_{n-2},\hdots,x_0,1)$ and $j+2=(y_{n-2},\hdots,y_0,1)$ are odd integers, the last bit in their binary representations are $1$ and the Hamming distance between $j$ and $j+2$ is $l\in\{1,\hdots,n-1\}$ with $y_i=x_i\oplus 1$ at $l$ indices $i$, and otherwise $y_i=x_i.$

Now we define quantum gates that transform $\ket{j}_n=\ket{x1}_n$ to $\ket{j+2}_n=\ket{(x+1)1}_n$ and vice-versa in an $n$-qubit system, when $j$ is odd. We exploit the last bit of $j$ and $j+1$ being $1$ in their binary representations. Indeed, a quantum gate $U_x$ which performs this task is given by \begin{equation}\label{eqn:gateU}
    U_x=\begin{cases}
        X_{n(n-1)} \,\, \mbox{if} \,\, x \,\, \mbox{is even} \\
        \prod_{j\in \mathcal{I}} X_{n(n-i+1)} \,\, \mbox{if} \,\, x \,\, \mbox{is odd,}
    \end{cases}
\end{equation} where $\mathcal{I}=\{i: y_i\neq x_i\}$, $\ket{j}_n=\ket{x1}_n=\ket{x_{n-2},\hdots,x_0,1}_n$ and $\ket{j+2}_n=\ket{(x+1)1}_n=\ket{y_{n-2},\hdots,y_0,1}_n$ with $x=(x_{n-2},\hdots,x_0),$ $x+1=(y_{n-2},\hdots,y_0),$ $x_j,y_j\in\{0,1\},$ $j=0,\hdots,n-2.$ Here $X_{kl}$ denotes the CNOT gate with $k$-th qubit as control and $l$-th qubit as the target qubit. Obviously, $|\mathcal{I}|$ equals the Hamming distance between $x$ and $x+1.$ For example, in the $3$-qubit system, if $j=5=(101)$ and hence $j+2=7=(111)$ then $U_x=X_{32}$, whereas $U_x=X_{31}X_{32}$ if $j=3=(011)$ and $j+2=5=(101)$ such that $U_x\ket{j}_3=\ket{j+2}_3$ and $U_x\ket{j+2}_3=\ket{j}_3.$ Note that the $(n-i+1)$-th qubit in $\ket{x1}_n$ is given by $x_{n-i}$ in  $x=(x_{n-2},\hdots,x_0).$

We emphasis here that the unitary gate $U_x$ given by equation (\ref{eqn:gateU}) not only non-trivially acts on $\ket{j}_n$ and $\ket{j+1}_n$ but also on other basis states of the $n$-qubit system, but in what follows we will see that those actions will be nullified by repeating the action of $U_x$ again towards the circuit representation of $s_j=(j,j+1),$ as shown in the following theorem.

\begin{theorem}
 Suppose $j=(x,1)$ be the $n$-bit representation of an odd integer $j\in\{0,1,\hdots,2^n-1\},$ $x=(x_{n-2},\hdots,x_0)\in\{0,1\}^{n-1}$ and $x\neq \boldsymbol{1}_{n-1}.$ Then the quantum circuit for the adjacent transposition $s_j=(j,j+1)$ for the set $\{0,1,\hdots,2^n-2\}$ is given by the unitary gate $U_xT^{(n)}_{x+1}U_x,$ where $U_x$ is given by equation (\ref{eqn:gateU}) and $T^{(n)}_{x+1}$ is the $n$-qubit quantum (generalized) Toffoli gate corresponding to the integer $x+1\in\{1,\hdots,2^{n-1}-1\}.$      
\end{theorem}

\pf Note that the action of $U_x$ on a quantum state $\ket{\psi}_n=\sum_{j=0}^{2^n-1} a_j\ket{j}_n$ makes the coefficients corresponding to the basis elements $\ket{j}_n$ and $\ket{j+2}_n$ are interchanged. Indeed, the coefficients corresponding to the terminal nodes $\ket{j+1}_n=\ket{(x+1)0}_n$ and $\ket{j+2}_n=\ket{(x+1)1}_n$ in $\QT_n$ are given by $a_{j+1}$ and $a_{j}$ respectively, after application of $U_x$ to $\ket{\psi}_n.$ Next we apply the quantum (generalized) Toffoli gate $T^{(n)}_{x+1}$ on $U_x\ket{\psi}_n,$ where $\ket{j+1}_n=\ket{(x+1)0}_n$ and $\ket{j+2}_n=\ket{(x+1)1}_n$ are the terminal nodes that stem from the $(n-1)$-qubit state $\ket{x+1}_{n-1}$ in $\QT_n.$ 

Then observe that $T^{(n)}_{x+1}$ acts non-trivially only on the basis states $\ket{j+1}_n=\ket{(x+1)0}_n$ and $\ket{j+2}_n=\ket{(x+1)1}_n$, and since it represents the transposition $(j+1,j+2)$ (from Section \ref{sec:cToffoli}) for the set $\{0,1,\hdots,2^n-1\},$ it's actions only interchanges the coefficients corresponding to the basis states $\ket{j+1}_n$ and $\ket{j+2}_n$ which are $a_{j+1}$ and $a_{j}$ in $U_x\ket{\psi}_n.$ Thus the coefficients of $\ket{j}_n$, $\ket{j+1}_n$ and $\ket{j+2}$ are given by $a_{j+2},$ $a_{j}$, and $a_{j+1}$ respectively. Finally, applying $U_x$ on $T^{(n)}U_x\ket{\psi}_n,$ the coefficients corresponding to the basis states $\ket{j}_n$ and $\ket{j+2}_n$ interchange, and the other coefficients in $\ket{\psi}_n$ that were replaced due to the action of $U_x$ at the first step, regains its positions. Hence, finally we have \begin{eqnarray}  \ket{\psi}_n = \sum_{i=0}^{2^n-1}a_i\ket{i}_n &\boldsymbol{\mapsto}& \sum_{i=0}^{j-1} a_i\ket{i}_n + a_{j+1}\ket{j}_n + a_j\ket{j+1}_n \nonumber \\ && +\sum_{i=j+2}^{2^n-1} a_i\ket{i}_n \nonumber \\ &=& U_xT^{(n)}_{x+1}U_x\ket{\psi}_n,\end{eqnarray} which implements $s_j=(j,j+1),$ where $j=(x,1)$ and $j+1=(x+1,0)$ are the binary representation of $j$ and $j+1$ respectively, $x\in\{0,1\}^{n-1}.$ This completes the proof. \hfill{$\square$}

\usetikzlibrary {arrows.meta}
\begin{figure}[htbp]%{$t$\textwidth}
				\centering
				\begin{tikzpicture}
     %\draw (0,0) node[draw,circle] {$t$};
     %\node[draw,circle] at (0,0) {\tb{t}};
				\draw [fill] (-0.5, 0) circle [radius = .1];
                \draw [fill] (1.5, 0) circle [radius = .1];
				\node [below] at (-0.5, 0) {$\ket{10}_2$};
                \node [below] at (1.5, 0) {$\ket{11}_2$};
				\draw [fill] (-1, 1) circle [radius = .1];
                \draw [fill] (0, 1) circle [radius = .1];
                \draw [fill] (1, 1) circle [radius = .1];
                \draw [fill] (2, 1) circle [radius = .1];
				\node [above] at (-1, 1) {$\substack{\tb{a_4} \\\ket{100}_3}$};
                \node [above] at (0, 1) {$\substack{\tb{a_5} \\ \ket{10\tre{1}}_3}$};
                \node [above] at (1, 1) {$\substack{\tb{a_6} \\ \ket{110}_3}$};
                \node [above] at (2, 1) {$\substack{\tb{a_7} \\ \ket{11\tre{1}}_3}$};
            \draw (-0.5,0) -- (-1,1);
            \draw (-0.5,0) -- (0,1);
            \draw (1.5,0) -- (1,1);
            \draw (1.5,0) -- (2,1);
            \draw [arrows = {-Stealth[scale width=2]}] (2.5,0.5) -- (3.5,0.5);
            \node [above] at (2.9, 0.5) {$X_{32}$};
			%\draw (-2,1) -- (0, 0);
          \draw [fill] (4.5, 0) circle [radius = .1];
                \draw [fill] (6.5, 0) circle [radius = .1];
				\node [below] at (4.5, 0) {$\ket{10}_2$};
                \node [below] at (6.5, 0) {$\ket{11}_2$};
				\draw [fill] (4, 1) circle [radius = .1];
                \draw [fill] (5, 1) circle [radius = .1];
                \draw [fill] (6, 1) circle [radius = .1];
                \draw [fill] (7, 1) circle [radius = .1];
				\node [above] at (4, 1) {$\substack{\tb{a_4} \\\ket{100}_3}$};
                \node [above] at (5, 1) {$\substack{\tb{a_7} \\ \ket{10\tre{1}}_3}$};
                \node [above] at (6, 1) {$\substack{\tb{a_6} \\ \ket{110}_3}$};
                \node [above] at (7, 1) {$\substack{\tb{a_5} \\ \ket{11\tre{1}}_3}$};
            \draw (4.5,0) -- (4,1);
            \draw (4.5,0) -- (5,1);
            \draw (6.5,0) -- (6,1);
            \draw (6.5,0) -- (7,1);
            \draw [red, <.<->.>] (5,1) to [bend right] (7,1);
            \draw [arrows = {-Stealth[scale width=2]}] (5.5,-0.5) -- (5.5,-1.5);
            \node [above] at (5.9, -1.2) {$T_{11}$};
            %\draw (-2,1) -- (0, 0);
          \draw [fill] (4.5, -3) circle [radius = .1];
                \draw [fill] (6.5, -3) circle [radius = .1];
				\node [below] at (4.5, -3) {$\ket{10}_2$};
                \node [below] at (6.5, -3) {$\ket{11}_2$};
				\draw [fill] (4, -2) circle [radius = .1];
                \draw [fill] (5, -2) circle [radius = .1];
                \draw [fill] (6, -2) circle [radius = .1];
                \draw [fill] (7, -2) circle [radius = .1];
				\node [above] at (4, -2) {$\substack{\tb{a_4} \\\ket{100}_3}$};
                \node [above] at (5, -2) {$\substack{\tb{a_7} \\ \ket{10\tre{1}}_3}$};
                \node [above] at (6, -2) {$\substack{\tb{a_5} \\ \ket{110}_3}$};
                \node [above] at (7, -2) {$\substack{\tb{a_6} \\ \ket{11\tre{1}}_3}$};
            \draw (4.5,-3) -- (4,-2);
            \draw (4.5,-3) -- (5,-2);
            \draw (6.5,-3) -- (6,-2);
            \draw (6.5,-3) -- (7,-2);
             \draw [red, <.<->.>] (6,-2) to [bend right] (7,-2);
            \draw [arrows = {-Stealth[scale width=2]}] (3.5,-2.5) -- (2.5,-2.5);
            \node [above] at (2.9, -2.5) {$X_{32}$};
            \draw [fill] (-0.5, -3) circle [radius = .1];
                \draw [fill] (1.5, -3) circle [radius = .1];
				\node [below] at (-0.5, -3) {$\ket{10}_2$};
                \node [below] at (1.5, -3) {$\ket{11}_2$};
				\draw [fill] (-1, -2) circle [radius = .1];
                \draw [fill] (0, -2) circle [radius = .1];
                \draw [fill] (1, -2) circle [radius = .1];
                \draw [fill] (2, -2) circle [radius = .1];
				\node [above] at (-1, -2) {$\substack{\tb{a_4} \\\ket{100}_3}$};
                \node [above] at (0, -2) {$\substack{\tb{a_6} \\ \ket{10\tre{1}}_3}$};
                \node [above] at (1, -2) {$\substack{\tb{a_5} \\ \ket{110}_3}$};
                \node [above] at (2, -2) {$\substack{\tb{a_7} \\ \ket{11\tre{1}}_3}$};
            \draw (-0.5,-3) -- (-1,-2);
            \draw (-0.5,-3) -- (0,-2);
            \draw (1.5,-3) -- (2,-2);
            \draw (1.5,-3) -- (1,-2);
             \draw [red, <.<->.>] (0,-2) to [bend right] (2,-2);
			\end{tikzpicture}
				\caption{Construction of quantum circuit for the transposition $(5,6)=X_{32}T_{11}X_{32}$ in $3$-qubit system.}
				\label{fig:tree3}
\end{figure}
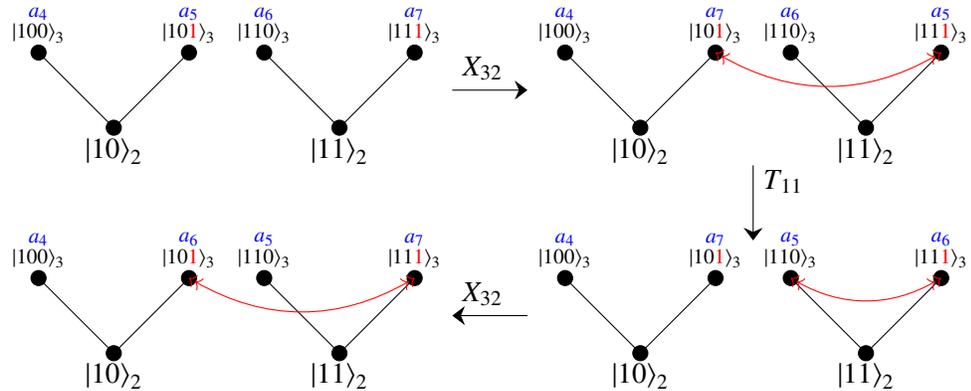

\begin{figure}[htbp]
    \centering
     \subfigure [\centering ]{{\includegraphics[width=0.15\textwidth]{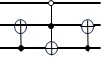} }}%
     %\caption{}
    \qquad
     \subfigure [\centering ]{{\includegraphics[width=0.15\textwidth]{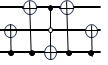} }}%
    \qquad
    \subfigure [\centering ]{{\includegraphics[width=0.15\textwidth]{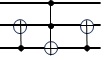} }}%
    %\qquad
    %\subfigure [\centering ]{{\includegraphics[width=0.42\textwidth]{T00.jpg} }}%
    % \qquad
   % \subfloat [\centering ]{{\includegraphics[width=0.35\textwidth]{lognormal-dist.png} }}%
\centering \caption {Quantum circuits for (odd,even) transpositions (a) $(1,2)$, (b) $(3,4)$, (c) $(5,6).$}
    \label{fig:oegates}
\end{figure}

In Figure \ref{fig:tree3}, we illustration the quantum circuit formation of the transposition $(5,6)$ in a $3$-qubit system. In Figure \ref{fig:oegates} we exhibit the quantum circuits for transpositions $s_j=(j,j+1)$, $j$ is odd for $3$-qubit system.

\subsection{Circuit complexity of adjacent transpositions}

Since a generalized Toffoli gate $T^{(n)}_x,$ $x\in\{0,1,\hdots, 2^{n-1}-1\}$ directly implements an adjacent transposition $s_j$ when $j$ is even (including $j=0$), the number of quantum gates needed to implement such a $s_j$ is at most $2(n-1)$ $X$ gates and one standard Toffoli gate, as described in Algorithm \ref{Alg:gTgate}. To be explicit, the number of $X$ gates required is $2k,$ when $k$ is the number of $0$s in the $(n-1)$-bit representation of $x.$ For circuit representation of $s_j$, when $j$ is odd, it requires a generalized Toffoli gate $T^{(n)}_x,$ where $j=(x,1)$ is the binary representation of $j$ with $x=(x_{n-2},\hdots,x_0).$ It also requires $2$ CNOT gates if $x$ is even or $2h$ CNOT gates when $x$ is odd, and $h$ is the Hamming distance between the $(n-1)$-bit binary representations of $x$ and $x+1.$ 

%Finally, since any permutation in $\mathcal{S}_N$, $N=2^n$ can be expressed as a product of at most $N(N-1)/2$ adjacent transpositions, we can calculate the number of $X$ gates, CNOT gates and the standard Toffoli gates to be used for circuit implementation of any permutation through the decomposition of adjacent transpositions as described in Algorithm \ref{algoPerp}.

\section{Quantum circuit model for random sampling of permutations}\label{Sec:qcforperm}

Building upon the classical random sampling procedure for permutations on $N$ elements, as outlined in Section~\ref{Sec:classical}, and utilizing the quantum circuit framework based on adjacent transpositions introduced in Section~\ref{sec:qcAT}, we propose a corresponding quantum circuit model. This model comprises two registers: a primary register of 
$\lceil\log_2N\rceil$-qubits for implementing permutations, and an ancillary register consisting of $N-1$ qudit states, of dimension 
$2\leq d\leq N$. We then present a quantum circuit design that realizes any specific permutation array directly on the primary register of $\lceil\log_2N\rceil$ qubits  without requiring ancillary states. Finally, we analyze and compare the asymptotic resource requirements, including the number of qubits and quantum gates, for both implementations. We denote a $d$ dimensional quantum state as $\ket{\psi}_d.$

%Indeed, the algorithm is a quantum implementation of the Algorithm \ref{algo1} via quantum circuits with the use of ancillary quantum states. The circuit model consists of two registers: the In this section 

%\subsection{Quantum circuit for generation of random permutation following uniform distribution}

We demonstrate the construction of the circuit model for the case $N = 2^n$, utilizing a primary register of $n$ qubits. A similar construction extends naturally to the general case $N \neq 2^n$, where the primary register consists of $\lceil \log_2 N \rceil$ qubits. First recall from Algorithm  \ref{algo1} that, in order to generation a random uniformly distributed permutation from $\mathcal{S}_N,$ $N\geq 3$, we need to pick uniformly randomly a permutation from the set $\Pi_k,$ $0\leq k\leq N-2$ for a given $N.$ For $n$-qubit system, $N=2^n.$ In the proposed circuit model of the quantum algorithm, we associate an ancillary quantum state of dimension $k+2$ for each $\Pi_k$ in order to pick an element from $\Pi_k$ in the main circuit. 

Recall from equation (\ref{eqn:pik}) that $$ \Pi_k=\left\{I,s_k,s_{k}s_{k-1}, s_ks_{k-1}s_{k-2},\hdots,  s_ks_{k-1}s_{k-2}\hdots s_0\right\},$$ for $0\leq k\leq N-2,$ where $s_j=(j,j+1),$ the adjacent transposition, and every element in $\mathcal{S}_N$ can be expressed as product of elements from $\Pi_k.$ Considering $\Pi_k$ as an ordered set, we define a $(k+2)$-dimensional quantum state belonging to the Hilbert space $\C^{k+2}$,  
\begin{equation}
    \ket{\Pi_k}_{k+2}=\frac{1}{\sqrt{k+2}}\left(\sum_{j=0}^{k+1} \ket{j}_{k+2}\right).
\end{equation} Here $\{\ket{j}_{k+2} : 0\leq j\leq k+2\}$ denotes the canonical basis of $\C^{k+2}.$ Then we introduce a controlled-$\Pi_k[j]$ gate for the $j$-th element of $\Pi_k$ corresponding to $\ket{j}_{k+2}$ with $\Pi_k[0]=I_N$, for the ordered elements $\Pi_k[j]\in \Pi_k.$  

Thus for the main quantum circuit for sampling a random permutation, we need $N-1=2^n-1$ 
 ancillary quantum states $\ket{\Pi_k}_{k+2},$ $0\leq k\leq N-2.$ The total composite ancilla quantum state is given by 
 \begin{eqnarray*}
    && \otimes_{k=0}^{N-2} \ket{\Pi_k}_{k+2} \\
    &=& \ket{\Pi_0}_2\otimes \ket{\Pi_1}_3 \otimes \cdots \otimes \ket{\Pi_{N-2}}_{N} \\
     &=& \frac{1}{\sqrt{2}}\left(\ket{0}_2+\ket{1}_2\right) \otimes \frac{1}{\sqrt{3}} \left(\ket{0}_3+\ket{1}_3+\ket{2}_3\right) \otimes \cdots \\
     && \hfill{\otimes \frac{1}{\sqrt{N}} \left(\ket{0}_N+\ket{1}_N+\cdots +\ket{N-1}_N\right)} \\
     &=& \frac{1}{\sqrt{N!}} \sum_{j=0}^{N!-1} \ket{j}_{23\cdots N},
 \end{eqnarray*} where $\{\ket{j}_{23\cdots N} : 0\leq j\leq N!-1\}$ denotes the canonical basis of the Hilbert space $\C^2\otimes \C^3\otimes \cdots \otimes \C^{N}.$

 Now performing a simultaneous quantum measurement to all the ancillary quantum states $\ket{\Pi_k}_{k+2},$ $0\leq k\leq N-2$ with respect to computation basis of their respective Hilbert spaces, we obtain a random permutation as a product of adjacent permutations from $\Pi_k$ that correspond to the measurement outcome of the basis state $\ket{j}_{23\cdots N}.$ This random permutation should be then acted on the $n$-qubit input state $\ket{\psi}_{N}$, whose probability amplitudes encode the array of $N=2^n$ elements to be permuted. 

%We illustrate the circuit model of the algorithm for $2$-qubit system first. For $2$-qubit system the quantum circuit for sampling a random permutation from $\mathcal{S}_{2^2}=\mathcal{S}_4$ is given by Figure \ref{fig:s4_rpqc}. Here the ancillary quantum states $\ket{\Pi_0}$ 

% \begin{figure}[htbp]
 %   \centering
  %   \includegraphics[width=0.70\textwidth]{s4_rpqc.jpg} 
%\centering \caption {Quantum circuit for sampling random permutation on $n$-qubit system}
 %   \label{fig:s4_rpqc}
%\end{figure}

\begin{figure}[htbp]
    \centering
     \includegraphics[width=0.50\textwidth]{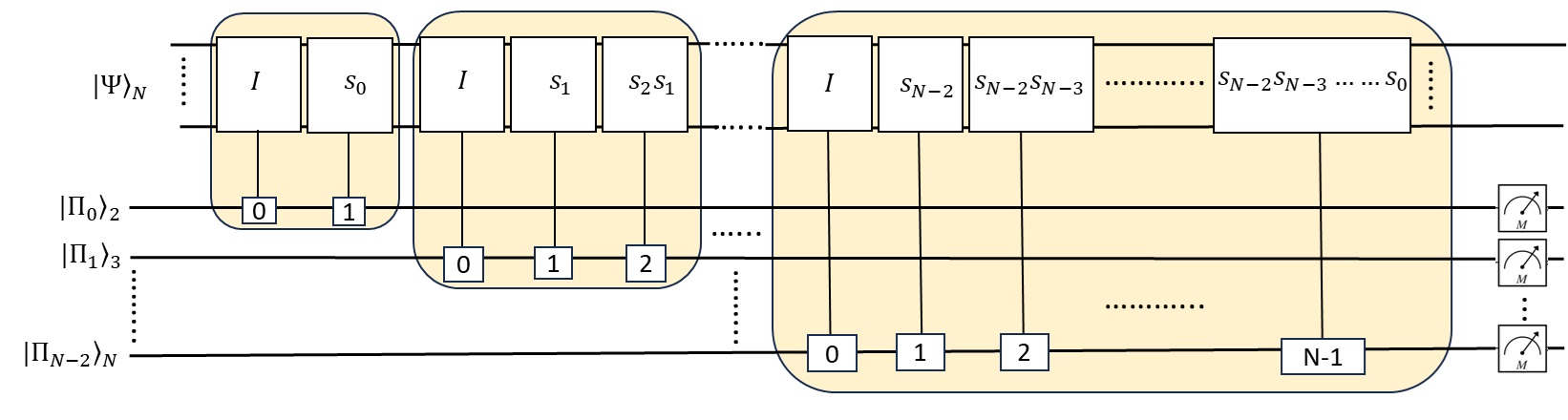} 
\centering \caption {Quantum circuit for sampling random permutation on $n$-qubit system. The orange box indicates the permutations from the ordered set $\Pi_k,$ $0\leq k\leq N-1.$ }
    \label{fig:sN_rpqc}
\end{figure}

Further, note that choosing the first $k-1$ ancillary quantum states $\ket{\Pi_i}_{i+2},$ $0\leq i\leq k-2$ for simultaneous measurements, it implements a random sampling from the symmetric group $\mathcal{S}_{k}$ for $k\geq 2,$ treating the permutations on $N$ elements while the permutation nontrivially acts only on the first $k$ elements. Thus the quantum circuit model can be employed to sample from any symmetric group $\mathcal{S}_k$ where $2\leq k\leq N=2^n$ on an $n$-qubit system. Besides, we show in Section \ref{Sec:corona} that a similar circuit can be defined for sampling from specific subsets of $\mathcal{S}_N.$ 

%\tre{The advantage of using the quantum circuit model for sampling permutations over the classical algorithms proposed in Section \ref{Sec:classical} is that, it does not involve performing any classical computation.} \tm{\tb{A distinctive feature of the quantum circuit model for sampling permutations, in contrast to the classical algorithms discussed in Section \ref{Sec:classical}, is its requirement of only $\log N = n$ qubits to encode the $N = 2^n$ elements in the input quantum state $\ket{\Psi}_N$. On the other hand, the space complexity of the Steinhaus-Johnson-Trotter algorithm is $O(N)$.} The product of adjacent transpositions is implemented by the controlled gates defined by the ancillary quantum states due to the mechanism of quantum circuit. However, it should be noted that the sampling of permutation from $\Pi_j$ or $K_j$ (as dictated by Algorithm \ref{algo1} and Algorithm \ref{algo12}) whose parallel in the quantum algorithm is the measurement of of high-dimensional states $\ket{\Pi_k}_{k+2}$ in the ancilla register. \tb{Note that the ancillary state of dimension $d$, where $2 \leq d \leq N$, can be realized using $\lceil \log_2 d \rceil$ qubits. The quantum resource requirements of the proposed model are discussed in Section~\ref{Sec:qresource}.}}

Note that the ancillary quantum states of dimension $2 \leq d \leq N$ in the proposed quantum circuit model can be prepared using $\lceil\log_2 d\rceil$ qubits. Consequently, the total number of qubits required for the ancillary register is given by $\sum_{d=2}^{N} \lceil\log_2 d\rceil < (N-1)\lceil\log_2 N\rceil$. Including the $\lceil\log_2 N\rceil$ qubits for the primary register, which implements the random permutation, the total number of qubits required is strictly less than $N\lceil\log_2 N\rceil$.

%\tb{Furthermore, according to \cite{shukla2024efficient}, the ancillary quantum state of dimension (sparsity) $d$ of a $\lceil\log_2d\rceil$-qubit system can be constructed using $O(\log_2 d)$ quantum gates. Summing over all $d = 2$ to $N$, the total gate count to prepare the ancillary register is $O(N\log_2 N)$. }

Furthermore, according to Theorem 1 of \cite{gleinig2021efficient}, the ancillary quantum state of dimension (sparsity) $d$ of a $\lceil\log_2d\rceil$-qubit system can be constructed using $O(d \log_2 d)$ one-qubit gates and $O(\lceil d \log_2 d \rceil)$ CNOT gates. Summing over all $d = 2$ to $N$, the total gate count to prepare the ancillary register is $O(N^2\log_2 N)$.

For the primary register, which uses $\lceil\log_2 N\rceil$ qubits, the implementation involves circuits for the elements of $\Pi_k$ for $0 \leq k \leq N-2$. Each $\Pi_k$ consists of product of $l_k$ adjacent transpositions, where $1 \leq l_k \leq k+2$. Thus $\Pi_k$ is formed by total $1+2+\hdots+(k+1)=(k+1)(k+2)/2$ adjacent transpositions, and hence the total number of adjacent transpositions in the primary register of circuit in Figure \ref{fig:sN_rpqc} is 
\[\sum_{k=0}^{N-2} \frac{(k+1)(k+2)}{2} < \frac{N^2(N-1)}{2}.
\]
Each adjacent transposition $s_j = (j, j+1)$ is synthesized using a generalized Toffoli gate if $j$ is even, and a generalized Toffoli gate along with up to $h$ CNOT gates if $j$ is odd, where $h < \lceil\log_2 N\rceil$ is the Hamming distance between $x$ and $x+1$, and $j = (x,1)$ is the binary encoding of $j$. Therefore, the overall gate complexity for the complete quantum circuit implementation of random sampling of permutations on $N$ elements is $O(N^3 \log_2 N)$. Observe that, upon measurement of the ancillary register, the primary register implements the sampled permutation using $O(N^2 \log_2 N)$ elementary gates, reflecting a distinctive feature of the proposed construction.

%Finally we compare the asymptotic scaling of  qubit and gate count of the existing quantum circuit models for random sampling of permutations and the proposed model in Table \ref{Tab:qs}. Recall that the existing sampling algorithms are proposed in the literature for generating uniform superposition of permutations to be applicable in different contexts. Besides, each proposal of quantum circuit is unique to the use of  specific gate set, such as controlled generalized Toffoli and CNOT gates in this paper. In \cite{barenco1997stabilization} the authors consider projection into the symmetrised subspace with the use of control SWAP gates and specialized multi-qubit gates acting on the ancilla register. In \cite{chiew2019graph}, the authors consider quantum algorithm for graph similarity, the circuit is based on specialized unitary operations and it used mixed radix numeral system to numbering permutations. In \cite{bartschi2020grover},   the circuit is proposed using controlled SWAP and controlled cyclic rotation gates for uniform superposition of permutation matrices.

Finally, we compare the asymptotic scaling of qubit and gate complexity for existing quantum circuit models for random sampling of permutations with the proposed model in Table~\ref{Tab:qs}. It is important to note that existing approaches are primarily designed for generating a uniform superposition over permutations, with applications spanning various computational tasks. Each of these models employs a distinct quantum gate set, making a direct comparison nuanced.

The proposed model in this paper leverages generalized Toffoli gates, CNOT gates and $X$ gates to synthesize adjacent transpositions, and  thereby enabling random sampling of permutations using elementary quantum gates. In contrast, the model in \cite{barenco1997stabilization} is proposed in the context of generating symmetrized subspace for stable quantum computation using controlled SWAP gates and specialized multi-qubit gates acting on an ancilla register. We note that the circuit model proposed in \cite{barenco1997stabilization} utilizes a primary register of $N$ qubits to implement permutations, assisted by $O(N^2)$ ancilla qubits. In contrast, the primary register in our proposed method comprises only $\lceil\log_2N\rceil$ qubits. Consequently, the controlled SWAP gates can generate only $\lceil\log_2N\rceil !$ permutations, which is a small subset of the full $N!$ permutations. However, by employing controlled generalized Toffoli gates and controlled CNOT gates, our approach enables the generation of a uniform superposition over all $N!$ permutations, while requiring a total of $O(N\log_2N)$ qubits. The approach in \cite{chiew2019graph}, designed for a quantum algorithm addressing graph similarity, employs specialized unitaries and mixed radix numeral system to encode permutations. Meanwhile, \cite{bartschi2020grover} presents a quantum circuit that employs controlled SWAP gates and controlled cyclic rotation gates to construct a uniform superposition over permutation matrices, aimed at solving combinatorial optimization problems such as the Traveling Salesperson Problem and Discrete Portfolio Rebalancing. Moreover, the proposed circuit model is specifically designed for applications involving the permutation of $N$ classical data points that are encoded into a $\lceil\log_2 N\rceil$-qubit quantum state $\ket{\Psi}_N$ via probability amplitudes in the primary register.

%Quantum methods that are not specifically designed for implementing and manipulating permutations on a $\lceil\log_2 N\rceil$-qubit register may not be applicable in scenarios where $N$ classical data points are encoded in such a register.

It should be noted that the use of elementary quantum gates in the proposed circuit model enhances its suitability for implementation on near-term quantum hardware. In contrast, existing models often rely on specialized unitaries, whose decomposition into elementary gates may pose significant challenges for practical realization.

\begin{table}[]
    \centering
    \begin{tabular}{|c|c|c|c|}
    \hline
        Ref.  & qubit count  & gate count  \\ \hline\hline
        \cite{barenco1997stabilization}  & $O(N^2)$ & $O(N^2)$ \\
        %\hline
        \cite{chiew2019graph} & $O(N\log_2 N)$ & $O(N^3(\log_2N)^2)$ \\
        \cite{bartschi2020grover} & $O(N^2)$ & $O(N^3)$ \\
        This paper & $O(N \log_2 N)$ & $O(N^3\log_2 N)$ \\
        \hline
    \end{tabular}
    \caption{Comparison of asymptotic scaling of qubit count and gate count for different quantum algorithms for random sampling of permutations}
    \label{Tab:qs}
\end{table}

%However, these measurement steps can be performed by a classical sampling from the set $\{0,1,\hdots,k+1\}$ replacing the quantum measurement of the states $\ket{\Pi_k}_{k+2}$, where an ancillary state can be chosen as $\ket{j}_{k+2}$ to define the control gate when the classical sampling outcome is $j\in \{0,1,\hdots,k+1\}.$ 

\subsection{Quantum circuit implementation of permutation arrays} 
%\tb{Given a permutation array $\pi$ on $N$ elements, using Algorithm \ref{algo14} it can be expressed as a product of adjacent transpositions and hence using the quantum gate synthesis of adjacent transpositions we can  design a quantum circuit model without the use of any ancillary quantum states for implementation of a specific permutation of $N$ data points using a register of $\lceil\log_2N\rceil$ qubits.}

Given a permutation array $\pi$ on $N$ elements, Algorithm~\ref{algo14} expresses it as a product of adjacent transpositions. Leveraging the quantum gate synthesis for such transpositions, the proposed quantum circuit model given by Figure \ref{fig:sN_rpqc} enables the implementation of any specific permutation using only the primary register of $\lceil \log_2 N \rceil$ qubits, without the need for ancillary quantum states. The overall gate complexity of this construction is $O(N^2 \log_2 N)$.

In particular, two-qubit SWAP gates in an $n$-qubit system ($n \geq 2$) represents a permutation matrix of order $2^n.$ It play a pivotal role in various quantum algorithms, such as the SWAP test used for estimating the inner product between quantum states \cite{barenco1997stabilization}. A SWAP gate exchanges the quantum states of two specified qubits, thereby permuting certain computational basis elements and effectively implementing a permutation over the $2^n$-dimensional Hilbert space. In what follows, we discuss the permutation array induced by a SWAP gate, denoted as $\mbox{SWAP}_{i,j},$ acting on qubits indexed by $i$ and $j$ in an $n$-qubit system, where $0 \leq i, j \leq n - 1$. Consequently, a quantum circuit implementation of SWAP gates can be constructed using generalized Toffoli, CNOT, and $X$ gates.

%\tb{The two-qubit SWAP gates in an $n$-qubit system, $n\geq 2$ plays a pivotal role in different quantum algorithms such as performing SWAP test, which estimates the inner product between two quantum states. In \cite{barenco1997stabilization}, SWAP gates are used as a primary gates for implementing superposition of permutations. Mathematically, a SWP gate, which exchanges the states of the operational two qubits, permute certain basis elements in its outcome and it represents a permutation of the $2^n$ basis elements. Below we discuss the permutation array corresponding to a SWAP gate which acts on the qubits indexed by $i$ and $j$ in an $n$-qubit system, $0\leq i,j\leq n-1.$ As a consequence, we provide a quantum circuit implementation of SWAP gates through generalized Toffoili gate and CNOT gate. }

%{We denote a SWAP gate as $\mbox{SWAP}_{i,j}$ that exchanges the $i$-th and $j$-th qubit states in an  $n$-qubit circuit, $0\leq i,j\leq n-1.$ In what follows, we derive the permutation corresponding to $\mbox{SWAP}_{i,j}$ acting on the $n$-qubit Hilbert space. Then a circuit representation of a SWAP gate using generalized Toffoli gate and CNOT can be obtained by expressing the permutation as a product of adjacent transpositions using Algorithm \ref{algo14}, and then implementing the adjacent transpositions as described in Section \ref{sec:qcAT}. }

Consider the $n$-qubit computational basis states as defined in Equation~(\ref{eqn:basis}). When the SWAP gate $\mathrm{SWAP}_{i,j}$ acts on the $q$-th basis state $\ket{q_{n-1} \cdots q_{j+1} q_j q_{j-1} \cdots q_{i+1} q_i q_{i-1} \cdots q_0}$, where each $q_l \in \{0,1\}$ for $0 \leq l \leq n-1$, it exchanges the values of qubits $q_i$ and $q_j$. This operation effectively swaps the positions of elements in the permutation array corresponding to basis states where $q_i \neq q_j$. The resulting permutation array induced by $\mathrm{SWAP}_{i,j}$ can therefore be derived from the identity permutation by interchanging the indices
\[
\sum_{\substack{l=j+1}}^{n-1} q_l 2^l + \sum_{l=0}^{j-1} q_l 2^l \quad \text{and} \quad \sum_{\substack{l=i+1}}^{n-1} q_l 2^l + \sum_{l=0}^{i-1} q_l 2^l,
\]
while keeping all other entries unchanged. Since the swap affects only those basis states for which $q_i \neq q_j$, the total number of such interchanges is $2^{n-2}$, corresponding to all bit strings of length $n-2$ formed by fixing bits $\{q_l\}_{l \neq i,j}$.

\subsection{Quantum two-sample randomization test for classical data}

In this section, we introduce a quantum analogue of the classical Randomization Test (RT) in \textit{randomization model} used in nonparametric statistics for comparing two populations with minimal assumptions, see \cite{fisher1939comparison} \cite{lehmann2006nonparametrics} \cite{higgins2004introduction}. In particular, the  two-sample RT is often demonstrated in the context of evaluation of a new treatment for post-surgical recovery against a standard treatment by comparing the recovery times  of patients undergoing each treatment \cite{ernst2004permutation}. If $N$ subjects are available for the study, the objects are divided into two sets randomly to receive the new treatment. Suppose $K$ and $N-K$ objects are selected in first set and the second set, respectively. Then the null hypothesis and an alternative hypothesis for the test is defined by  $H_0$: \textit{There is no difference between the treatments}, and $H_1$: \textit{The new treatment decreases recovery times}, respectively. If the recovery times for the standard and new treatments are given by $X_1,\hdots, X_K$ and $Y_1,\hdots, Y_{N-K}$ respectively, then a usual measure to calculate the difference between the treatments is given by the test statistic $T=\overline{X}-\overline{Y},$ i.e. the difference of the means of the recovery times. It should be noted that the recovery times are not random but the assignment of the objects to the treatments is random. Therefore, the  probability distribution of $T$ can be given by the randomization of the available subjects to the treatments. Moreover, $p$-value of the test of $H_0$ is calculated as the probability of getting a test statistic as extreme as, or more extreme than (in favor of $H_1$), the observed test statistic $t^*$. Since there are ${N\choose K}$ randomization that are equally likely under $H_0$, the $p$-value is given by $$p = P(T\leq t^*|H_0)=\frac{\sum_{i=1}^{{N\choose K}}\mathbb{I}(t_i\leq t^*)}{{N\choose K}},$$ where $t_i$ is the value of the test statistic $T$  for
the $i$-th randomization and $\mathbb{I} (\cdot)$ is the indicator function. Obviously, the time complexity of calculating the means of ${N\choose K}$ two-samples is $O(N\cdot {N\choose K}).$

\begin{figure}[htbp]
    \centering
     \includegraphics[width=0.50\textwidth]{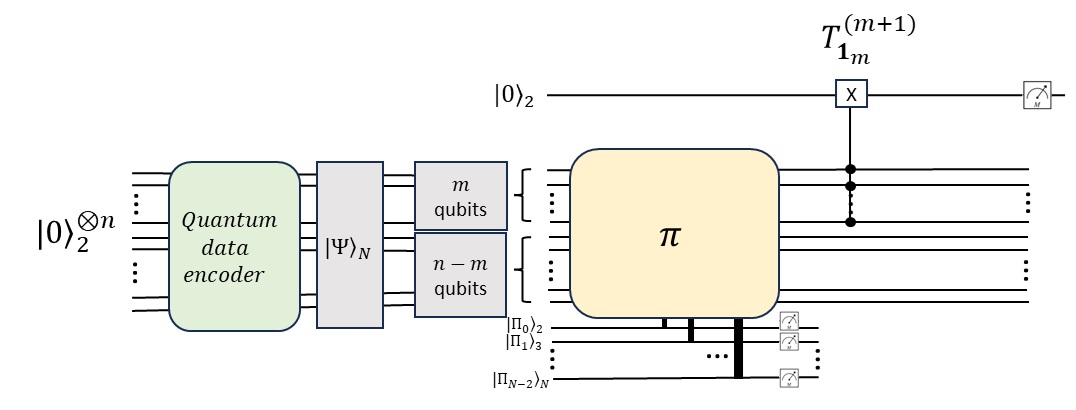} 
\centering \caption {Quantum circuit implementation of randomization test}
    \label{fig:s4_rpqc}
\end{figure}

In what follows, we propose to perform the randomization test using a quantum algorithm that can give an advantage to speed up the execution of the test. First we encode the given classical $2^n=N$ data points into an $n$-qubit quantum state. Then we perform a quantum measurement to an ancillary qubit which provides the means of two samples consisting of $K=2^{n-m}$ and $N-K$ data points for a choice of $1\leq m< n.$ Given a collection of $N=2^n$ positive data points $a_j\geq 0$, the Algorithm \ref{Alg:two-sampleRT} describes a quantum circuit model based algorithm for performing the two-sample randomization test. The quantum circuit which implements the algorithm is given by Figure \ref{fig:s4_rpqc}.

First recall that, any $n$-qubit state can be written as $\ket{\psi}_{N}=\sum_{j=0}^{N-1} x_j\ket{j}_{N},$ $N=2^n$ where $\{\ket{j}_{N} : 0\leq j\leq N-1\}$ denotes the canonical computational basis for the $n$-qubit system. Then for a permutation gate $\pi\in \mathcal{S}_{N},$ we have \begin{equation}\label{eqn:qpi}
    \pi\ket{\psi}_{N}=\sum_{j=0}^{N-1} x_{\pi(j)}\ket{j}_{N}.
\end{equation}  It is needless to say that an $n$-qubit register carries a  probability distribution. Besides, from equation (\ref{eqn:qpi}) it follows that when a permutation $\pi$ acts on an $n$-qubit state, it preserves the probability distribution described by the quantum measurement wrt the canonical basis, since $\{|x_j|^2 : 0\leq j\leq N-1\}=\{|x_{\pi(j)}|^2 : 0\leq j\leq N-1\}.$

\begin{algorithm}
  \caption{Quantum algorithm for two-sample randomization test}
  \begin{algorithmic}[1]
    \State Create a distribution $p_j,$ $0\leq j\leq 2^n -1$ where $p_j=a_j/(\sum_{j=0}^{2^n-1}a_j)$ and implement it through a quantum data loader with an $n$-qubit quantum state $$\ket{\Psi}_N=\sum_{j=0}^{2^n-1} \sqrt{p_j}\ket{j}_N.$$
    %\algrule
    \State Apply a random permutation $\pi$ on $\ket{\Psi}_N$ through the aforementioned procedure by performing simultaneous quantum measurements on the ancillary quantum states $\ket{\Pi_l}_{l+2},$ $0\leq l\leq N.$
    %\algrule[5pt]
    \State Record the measurement outcome of ancillary quantum state $\ket{\Pi_l}_{l+2}$ for classical data analysis. 
    \State Then apply the $(m+1)$-qubit Toffoli gate $T_{\boldsymbol{1}_m}^{(m+1)}$ with control qubits as the first $m$ qubits of the $n$-qubit register and the target qubit as the ancillary qubit $\ket{0}_2$ (the first qubit of the entire $(n+1)$-qubit register, which we call the `First qubit').
    \State Perform a quantum measurement with respect to the Pauli $Z$ matrix on the First qubit, .
    \State Record the outcome of the First qubit measurement.
    \State Repeat the steps $2$ to $6$. 
  \end{algorithmic}
  \label{Alg:two-sampleRT}
\end{algorithm}

%\begin{enumerate}
 %   \item Create a distribution $p_j,$ $0\leq j\leq 2^n -1$ where $p_j=a_j/(\sum_{j=0}^{2^n-1}a_j)$ and define an $n$-qubit quantum state $$\ket{\Psi}=\sum_{j=0}^{2^n-1} \sqrt{p_j}\ket{j}.$$
  %  \item Apply a random permutation $\pi$ on $\ket{\Psi}$ through the procedure described in Algorithm \ref{?} by performing simultaneous quantum measurements on the ancillary quantum states $\ket{\Pi_l}_{l+2},$ $0\leq l\leq N.$
   % \item Record the measurement outcome of ancillary quantum state $\ket{\Pi_l}_{l+2}$ for classical data analysis.  
   % \item Then apply the $(m+1)$-qubit Toffoli gate with control qubits as the first $m$ qubits of the $n$-qubit register and the target qubit as the ancillary qubit $\ket{0}_2$ (the last qubit of the entire $(n+1)$-qubit register)
   % \item Then perform a quantum measurement with respect to the Pauli $Z$ matrix on the last ancillary qubit 
   % \item Record the outcome of the last qubit measurement.  
   % \item Repeat the steps $2$ to $6$. 
%\end{enumerate}

{\bf Analysis of the algorithm:} One iteration of the algorithm would give us a pair of outcomes to keep in record for classical analysis of these outcomes for calculation of the means of the two samples of $K=2^{n-m}$ and $N-K=2^n-2^{n-m}$ data points. First observe that the permutation $\pi$ is obtained as $\prod_{s_l\in \Pi_l} s_l,$ where each $s_l$ is observed after measuring $\ket{\Pi_l}_{l+2},$ $0\leq l\leq N-2$ wrt the computational basis measurement. Next, observe that the Toffoli gate $T_{\boldsymbol{1}_m}^{(m+1)}$ with control qubits as the first $m$ qubits of the $n$-qubit register is acted on the quantum state 
\begin{eqnarray*} && \pi \ket{\Psi}_N \ket{0}_2 = \sum_{j=0}^{2^n-1} \sqrt{p_{\pi(j)}}\ket{j}_N\ket{0}_2 \\
&& = \sum_{\substack{j=(j_{n-1},\hdots,j_0) \\ j_l\in\{0,1\}}} \sqrt{p_{\pi(\sum_{l=0}^{n-1}j_l2^{l})}}\ket{j_{n-1}j_{n-2}\hdots j_0}_N\ket{0}_2.\end{eqnarray*} Then the output quantum state after applying the Toffoli gate is given by 
\begin{eqnarray}
 && T_{\boldsymbol{1}_m}^{(m+1)}\pi  \ket{\Psi}_N \ket{0}_2 \nonumber \\ &=&  \sum_{\substack{j_l\in\{0,1\}\\ (j_{m-1},\hdots,j_0)={\bf 1}_m}} \sqrt{p_{\pi(\sum_{l=0}^{n-1}j_l2^{l})}}\ket{j_{n-1}j_{n-2}\hdots j_{0}}_N\ket{1}_2  \nonumber  \\ && + \sum_{\substack{j_l\in\{0,1\}\\ (j_{m-1},\hdots,j_0)\neq {\bf 1}_m}} \sqrt{p_{\pi(\sum_{l=0}^{n-1}j_l2^{l})}}\ket{j_{n-1}j_{n-2}\hdots j_{0}}_N\ket{0}_2, \label{eqn:tgop}
\end{eqnarray} where ${\bf 1}_m$ is the all-one vector of dimension $m.$

It follows from the expression of the right-hand side of equation (\ref{eqn:tgop}) that there are $K=2^{n-m}$ quatum states with First qubit state $\ket{1}_2$, and $N-K=2^n-2^{n-m}$ quantum states with First qubit state $\ket{0}_2.$ Obviously, the the probability that $\ket{1}_2$ or $\ket{0}_2$ is obtained after $Z$-axis measurement to the First qubit is given by \begin{eqnarray}
   && p=P\left(\mbox{First qubit} = \ket{1}_2\right) = \sum_{\substack{j_l\in\{0,1\}\\ (j_{m-1},\hdots,j_0)={\bf 1}_m}} p_{\pi(\sum_{l=0}^{n-1}j_l2^{l})} \,\, \mbox{and} \nonumber \\
   &&  q=P\left(\mbox{First qubit} = \ket{0}_2\right) = 1-  p \nonumber \\ && \hfill{= \sum_{\substack{j_l\in\{0,1\}\\ (j_{m-1},\hdots,j_0)\neq {\bf 1}_m}} p_{\pi(\sum_{l=0}^{n-1}j_l2^{l})},} \label{eqn:zp}
\end{eqnarray} respectively.

Now note that corresponding to the control qubits, which are the first $m$ qubits of the $n$-qubit register, for the Toffoli gate $T_{\boldsymbol{1}_m}^{(n)}$, the $K$  basis elements of the $n$-qubit system in the rhs of equation (\ref{eqn:tgop}) are fixed whose corresponding coefficients determine $p$ and $q$ for any permutation $\pi$. These basis elements are given by $\ket{k},$ where \begin{equation}\label{eqn:k} k=\sum_{j=0}^{m-1} 2^j + \sum_{\substack{l=m \\ j_l\in\{0,1\}}}^{n-1} j_l2^{j_l},\end{equation} which we collect to denote the set $K$ for brevity. Each permutation assigns coefficients to these basis elements and there will be $K!\times (N-K)!$ permutations which will place the same set of coefficients in different permutations which will essentially be be used to compute $p$ and hence $q=1-p.$ While recording the permutation in each iteration, whether two different permutations, say $\pi$ and $\tau$ assign the same set of $K$ coefficients can be checked if and only if $\{\pi(j) | j\in K\}=\{\tau(j) | j\in K\}.$ Consequently, there are ${N\choose K}$ number of different sets of the coefficients which is equivalent to a partition of the permutation group $\mathcal{S}_{N}$ into ${N\choose K}$ classes, each of which contains $K!\times (N-K)!$ permutations. 

{\bf Classical processing of the data obtained from the quantim circuit:} Thus after generation of a random permutation $\pi$ through quantum circuits as a subroutine, it needs to be decided, which class it should belong to out of ${N\choose K}$ classes, which can be performed classically by determining the set $K_\pi=\{\pi(j) : j\in K\},$ where $K$ is composed of all $k$ given by equation (\ref{eqn:k}). The time complexity of obtaining the set $K_\pi$ for a $\pi$ is $O(1).$ Writing the set $K_\pi$ as an array $[K_\pi],$ the worst-case complexity of checking whether two arrays corresponding to two permutations $\pi, \tau$ obtained from the quantum random sampling method are identical is $O(K).$ Since there are ${N\choose K}$ sets of classes of permutations, and each randomly generated permutation belongs to one of these classes, upon quantum measurement of the First qubit, a classical register corresponding to each class will record the statistics of measurement outcomes either $0$ or $1$ according to the outcome $\ket{0}_2$ and $\ket{1}_2$ respectively. Thus the worst-case complexity up to this step is $O(K\cdot {N\choose K}).$ Now from the statistics of $0$ and $1$ in each class $C$ the probability value $p_c$ can be computed in $O(1)$ (which will be the same as described in equation (\ref{eqn:zp})). It should be note that the value of $p_c$ is obtained from a collection of at least $K!\times (N-K)!$ $0$ and $1,$ when all the $N!$ permutations (at least once) are obtained through the quantum circuit generation. 

%Upon the generation of $N!$ number of permutations through the randomization test quantum circuit (RTQC), we obtain the statistics of last qubit measurement outcome $\ket{0}_2$ and $\ket{1}_2$ for ${N\choose K}$ different two-samples choices. For a fixed class of permutations as discussed above, the algorithm produces a data of $K!\times (N-K)!$ qubits $\ket{0}_2$ and $\ket{1}_2,$ which ultimate estimates the probability $p$ and $q$ (as defined in equation (\ref{eqn:zp})). 

Finally, for a given class $C$ of the set of permutations, we have \begin{eqnarray*}
    p_c &=& \sum_{k\in K} p_k = \sum_{k\in K} \frac{a_k^c}{\sum_{j=0}^{2^n-1} a_j} = \frac{\sum_{k\in K}a_k^c}{\sum_{j=0}^{2^n-1} a_j} \\ &=& \mbox{Mean}\{a_k^c : k\in K\} \times \frac{K}{\sum_{j=0}^{2^n-1} a_j},
\end{eqnarray*} where $a_k^c$ is the set of data points belong to the class $C.$ Thus the mean of the sample data points corresponding to a class is obtained by multiplying $p$ with $\frac{\sum_{j=0}^{2^n-1} a_j}{K}.$ Thus the mean value for each sample can be estimated in $O(1)$ time from $p_c$. This concludes that the time complexity of processing the classical data obtained from the Algorithm \ref{Alg:two-sampleRT} is given by $O(K\cdot {N\choose K}),$ which shows a $O(2^m)$ improvement than the classical approach. For instance, if $m\approx n/2$ then a factor of $O(\sqrt{N})$ improvement can be observed.

The advantage of the proposed quantum algorithm is that, the mean values of the two samples for a class $C$ of permutations can be obtained from estimation of the probability values $p_c$ and $q_c=1-p_c,$ which are obtained by measurement of the ancillary First qubit for each random sampling of permutations. Finally, the $p$-value of the test can be calculated using these probability values to test the null hypothesis.

We have the following remarks about the proposed algorithm.
\begin{remark} 
   \begin{enumerate}
       \item The choice of the $m$ control qubits to apply the Toffoli gate can be chosen as any of the $m$ qubits from the $n$ qubits. Here we choose the first $m$ qubits.
       %\item The classical processing of the permutations to pick them the choice of the class that can be performed classically by comparing two vectors of length $K$ given by a permutation $\pi$ as $[\pi(k): k\in K]$
       \item Note that the algorithm need not be continued till all permutations are generated through the random sampling of permutation generation using the quantum circuit. When  $N$ is large, the number $K!\times (N-K)!$ is also a large number, and we do not need that many measurements to estimate the $p_c$ value for a class. Instead, a minimal number can be decided for the number of observations for each class to estimate $p_c$, and when it is reached the entire process can be stopped for classical processing of the data.  
   \end{enumerate} 
\end{remark}

From \cite{gleinig2021efficient}, we note that the state $\ket{\Psi}_N$ in Step~1 of Algorithm~\ref{Alg:two-sampleRT} can be prepared using $O(N\log_2 N)$ CNOT gates and $O(N\log_2 N)$ single-qubit gates. Consequently, the asymptotic scaling of the qubit count and gate count for Algorithm~\ref{Alg:two-sampleRT} are $O(N\log_2 N)$ and $O(N^3\log_2 N)$, respectively, which are the same as those of the quantum circuit model for random sampling of permutations described in Section~\ref{sec:qcAT}.

Finally, note that there are two primary challenges in realizing the proposed quantum circuit model on current quantum hardware. The first lies in encoding the $N$ elements into the probability amplitudes of the state $\ket{\Psi}_N$, and the second involves implementing adjacent transpositions using quantum gates within a noisy environment. The presence of noise in either component can significantly hinder the reliable execution of the circuit. A rigorous analysis of the model under realistic noise conditions is left for future investigation. Nevertheless, given that the circuit primarily relies on generalized Toffoli and CNOT gates, we anticipate that a successful implementation will be feasible either in a fault-tolerant quantum computing regime or on quantum architectures capable of high-precision realization of these gates.

\section{Random sampling from a specific set of permutations}\label{Sec:corona}

It is well known that a certain set of permutations is needed to perform various permutation tests \cite{ramdas2023permutation}. In this section, we propose a quantum measurement based procedure for sampling permutations from a desired set of permutations, which is decided by the choice of the ancillary quantum states for simultaneous measurement. We give a combinatorial perspective of the sampling method by introducing a nested corona product representation of symmetric groups.  

\subsection{Nested corona product graph representation of symmetric groups}

Various graph-theoretic structures have been proposed in the literature to represent symmetric groups, including Cayley graphs and the Sigma-Tau graph \cite{knuth2011art} \cite{sawada2019solving}. In this section, we introduce a graph representation of symmetric groups based on the corona product of graphs. We begin by recalling the definition of the corona product, originally introduced in the context of wreath products of groups, in particular, symmetric groups \cite{frucht1970corona}.

\begin{definition}\label{def:coronap}(Corona product of two graphs)
Let $G$ and $H$ be two graphs on $n$ and $k$ vertices respectively. Then the corona of $G$ and $H,$ denoted by $G\circ H$ is formed by taking one copy of $G$ and $n$ copies of $H$ such that $i$-th vertex of $G$ is joined by an edge of every vertex of the $i$-th copy of $H.$    
\end{definition}

\begin{figure}[htbp]  
\centering  
\subfigure[$G$]  
{  
\begin{tikzpicture}[scale=.8]  
\draw [fill] (0, 0) circle [radius = .1];
\draw [fill] (2, 0) circle [radius = .1]; 
 \draw (0,0) -- (2,0);
 \node [below] at (0, 0) {$a$};
                \node [below] at (2, 0) {$b$};
\end{tikzpicture}  
}  
% The only difference is here, where I have commented out an empty line.
\subfigure[$H$]  
{  
\begin{tikzpicture}[scale = 0.8]  
 \draw [fill] (3, 0) circle [radius = .1];
\draw [fill] (5, 0) circle [radius = .1];
\draw [fill] (4, 1) circle [radius = .1]; 
 \draw (3,0) -- (5,0);
\draw (3,0) -- (4,1);
\draw (4,1) -- (5,0);
 \node [below] at (3, 0) {$1$};
                \node [below] at (5, 0) {$2$};
                \node [above] at (4, 1) {$3$};
\end{tikzpicture}  
}
\subfigure[$G\circ H$]  
{  
\begin{tikzpicture}[scale = 0.8]  
\draw [fill] (6, 0) circle [radius = .1];
                \draw [fill] (8, 0) circle [radius = .1];
                \draw [fill] (10, 0) circle [radius = .1];
                \draw [fill] (12, 0) circle [radius = .1];
                \draw [fill] (7, 1) circle [radius = .1];
                 \draw [fill] (7, -1) circle [radius = .1];
                \draw [fill] (11, 1) circle [radius = .1];
                %\draw [fill] (8, 1) circle [radius = .1];
                \draw [fill] (11, -1) circle [radius = .1];
				\node [left] at (6, 0) {$3$};
                \node [above] at (8, 0) {$a$};
                \node [above] at (10, 0) {$b$};
                \node [right] at (12, 0) {$3$};
                \node [above] at (7, 1) {$1$};
                \node [above] at (11, 1) {$1$};
                \node [below] at (7, -1) {$2$};
                \node [below] at (11, -1) {$2$};
 \draw[dotted,thick] (6,0) -- (8,0);
            \draw (6,0) -- (7,1);
            \draw (6,0) -- (7,-1);
            \draw (7,1) -- (7,-1);
            \draw (8,0) -- (10,0);
            \draw[dotted,thick] (8,0) -- (7,1);
            \draw[dotted,thick] (8,0) -- (7,-1);
            \draw[dotted,thick] (10,0) -- (11,1);
            \draw[dotted,thick] (10,0) -- (11,-1);
            \draw (11,1) -- (11,-1);
			\draw[dotted,thick] (10,0) -- (12, 0);
   \draw (11,1) -- (12, 0);
   \draw (11,-1) -- (12, 0);
\end{tikzpicture}  
}
\caption{Corona product $G\circ H$ of two graphs $G$ and $H.$ The dotted edges correspond to the new edges due to corona product.}
\end{figure}
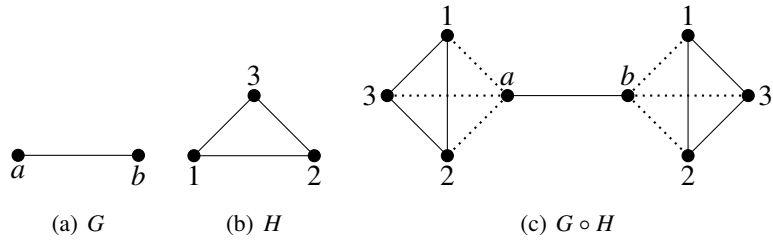

% \begin{figure}[htbp]
 %   \centering
  %   \includegraphics[width=0.60\textwidth]{coronapg.jpg} 
%\centering \caption {Corona product $G\circ H$ of two graphs $G$ and $H$}
 %   \label{fig:s4_rpqc}
%\end{figure}

Following Definition \ref{def:coronap}, if $V(G)$ and $V(H)$ denote the vertex sets of $G$ and $H$ respectively, then $G\circ H$ has $|V(G)| + (|V(G)|\times |V(H)|)$ number of vertices, whereas number of edges of $G\circ H$ is $|E(G)| + (|V(G)|\times |V(H)|) + (|V(G)| \times |E(G)|)=|E(G)|+ |V(G)|(|V(H)|+|E(G)|),$ where $E(X)$ denotes number of edges of the graph $X.$ Further, given a graph $G$ on $n$ vertices, and a collection of graphs $H_1,\hdots, H_n,$ the generalized corona graph, denoted by $G\circ \wedge_{i=1}^n H_i$ is defined as  taking one copies of $G,$ $H_1,\hdots,H_n$ join the $i$-th vertex of $G$ to all the vertices of $H_i$ by an edge \cite{laali2016spectra}. Further, given a graph $G^{(0)},$ the corona graphs are introduced in \cite{sharma2017structural} for the proposal of a large graph generative model by taking the corona product of $G^{(0)}$ iteratively. Indeed, for a given positive integer $m\geq 1,$ the corona graph $G^{(m)}$ is defined as $G^{(m)}=G^{(m-1)}\circ G^{(0)}.$ 

In what follows, we introduce \text{nested corona product} for a sequence of simple graphs $\{G_0, G_1,\hdots, G_{N-2}\}.$ By simple graph, we mean a graph with no loops and only one edge can exist between two vertices. 

\begin{definition}(Nested corona product of an ordered set of graphs)
Let $\{G_0, G_1,\hdots, G_{N-2}\}$ be a sequence of simple graphs. Then the nested corona product graph is defined as $$\circledwedge_{j=0}^{N-2} G_j := (\cdots ((G_0\circ G_1)\circ G_2)\cdots )\circ G_{N-2}.$$   
\end{definition}

Let $G^{\circledwedge N-1} :=\circledwedge_{j=0}^{N-2} G_j,$ $N\geq 3$ with $|V(G_j)|=n_j$ and $|E(G_j)|=m_j.$ Then there is a recurrence relation for the number of vertices for the nested corona product graphs given by \begin{equation}\label{eqn:rr}
    \left|V\left(G^{\circledwedge k}\right)\right| = \left|V\left(G^{\circledwedge k-1}\right)\right| + \left( \left|V\left(G^{\circledwedge k-1}\right)\right|\times \left|V\left(G_{k-1}\right)\right| \right), k\geq 2.
\end{equation}

%\tm{Thus $|V(G^{\circledwedge 2})|=|V(G_0\circ G_1)|=|V(G^{\circledwedge 1})| + (V(G^{\circledwedge 1})\times |V(G_1)|)= |V(G_0)|+(|V(G_0)|\times |V(G_1)|)=n_0+n_0n_1.$ Then $|V(G^{\circledwedge 3})|=|V((G_0\circ G_1)\circ G_2)|=|V(G^{\circledwedge 2})| + (V(G^{\circledwedge 2})\times |V(G_2)|)= n_0+n_0n_1 + (n_0+n_0n_1)n_2=n_0+n_0n_1+n_0n_2+n_0n_1n_2.$ This yields, $|V(G^{\circledwedge 4})|=|V(G_0\circ G_1\circ G_2\circ G_3)|=|V(G^{\circledwedge 3})| + (V(G^{\circledwedge 3})\times |V(G_3)|)= n_0+n_0n_1+n_0n_2+n_0n_1n_2+ (n_0+n_0n_1+n_0n_2+n_0n_1n_2)n_3=n_0+n_0n_1+n_0n_2+n_0n_3+n_0n_1n_2+n_0n_1n_3+n_0n_2n_3+n_0n_1n_2n_3.$ Consequently, $$|V(G^{\circledwedge 4})|=n_0\left[1+\sum_{l=1}^3 n_j + \sum_{\substack{l_1\neq l_2\\ l_1,l_2=1}}^3 n_{l_1}n_{l_2} + \prod_{l=1}^3n_l\right].$$}

%\tm{\begin{eqnarray} |V(G)| &=& n_0+n_0n_1+ (n_0+n_0n_1)n_2 + ((n_0+n_0n_1)n_2)n_3+\hdots  \,\,\mbox{and} \\  |E(G)| &=& m_0 + n_0(n_1+m_1) + (n_0+n_1)(n_2+m_2) + \hdots + (n_0+\hdots+n_{k-2})(n_{k-1}+m_{k-1}) \nonumber \\    &=& m_0+\sum_{j=0}^{k-2} (n_0+\hdots+n_j)(n_{j+1}+m_{j+1}).\end{eqnarray} }

 Before proceeding to derive the number of vertices and edges in a nested corona product graph, we recall the following from \cite{cox1997ideals}. 

The elementary symmetric polynomials for a given set of $n$ variables $x_1,\hdots, x_n$ are given by \begin{enumerate}
    \item $e_0(x_1,\hdots,x_n)=1$ (by convention)
    \item $e_1(x_1,\hdots,x_n)=\sum_{j=1} x_j$
    \item $e_2(x_1,\hdots,x_n)=\sum_{1\leq i< j\leq n} x_ix_j$
    \item $\vdots$
    \item $e_n(x_1,\hdots,x_n)=x_1x_2\cdots x_n.$
\end{enumerate} Further, \begin{equation}\label{eqn:symsum}
    \sum_{k=0}^n e_k(x_1,\hdots,x_n)=(1+x_1)(1+x_2)\cdots (1+x_n).
\end{equation}

\begin{theorem}\label{Thm:vencpg} $G^{\circledwedge N-1} :=\circledwedge_{j=0}^{N-2} G_j,$ $N\geq 3$ with $|V(G_j)|=n_j$ and $|E(G_j)|=m_j.$ Then \begin{eqnarray}
\left|V(G^{\circledwedge N-1} )\right| &=& n_0(1+n_1)(1+n_2)\cdots (1+n_{N-2}), \nonumber \\
\left|E(G^{\circledwedge N-1} )\right| 
    &=& m_0+\sum_{j=0}^{N-3} (n_0+\hdots+n_j)(n_{j+1}+m_{j+1}).\nonumber 
\end{eqnarray}    
\end{theorem}

\pf The proof for $\left|V(G^{\circledwedge N-1} )\right|$ follows from repeated application of the recurrence relation given by equation (\ref{eqn:rr}). Indeed, from the definition of nested corona product graph, it follows that \begin{eqnarray*}\left|V(G^{\circledwedge N-1} )\right| &=& n_0\left(1+\sum_{l=1}^{N-2} n_l + \sum_{\substack{l_1\neq l_2\\ l_1,l_2=1}}^{N-2} n_{l_1}n_{l_2} + \sum_{\substack{l_1\neq l_2\neq l_3\\ l_1,l_2,l_3=1}}^{N-2} n_{l_1}n_{l_2}n_{l_3} + \right.\\ 
&& \left. \hdots + \sum_{\substack{l_1\neq l_2\neq l_3,\hdots,\neq l_{N-3}\\ l_1,l_2,l_3,\hdots, l_{N-3}=1}}^{N-2} n_{l_1}n_{l_2}n_{l_3}\cdots n_{l_{N-3}} +\prod_{l=1}^{N-2} n_l\right).\end{eqnarray*} Then note that each term of the sum is a symmetric polynomial for $N-2$ variables, which are the number of vertices $n_1,\hdots, n_{N-2}.$  Then the result follows from equation (\ref{eqn:symsum}).

For the number of edges, the proof follows from the formation of the nested corona product graph, which is given by \begin{eqnarray*} \left|E(G^{\circledwedge N-1} )\right| &=& m_0 + n_0(n_1+m_1) + (n_0+n_1)(n_2+m_2) + \\ && \hdots + (n_0+\hdots+n_{N-3})(n_{k-1}+m_{N-2}).\end{eqnarray*}
This completes the proof. \hfill{$\square$}

Now we derive certain characteristics of $G^{\circledwedge N-1},$ $N\geq 3$ that will be used in sequel. First, observe that the number of copies of a $G_j,$ $2\leq j\leq N-2$ that gets attached during the formation of $G^{\circledwedge N-1}$ is given by $\left|V(G^{\circledwedge j})\right|=n_0(1+n_1)(1+n_2)\hdots (1+n_{j-1}),$ where $G^{\circledwedge j}= (\cdots((G_0\circ G_1)\circ G_2)\circ\cdots G_{j-1}).$ Obviously, number of copies of $G_1$ that gets attached is $n_0=|V(G_0)|.$ We denote the $i$-th copy of $G_j$ as $G_j^{(i)}$ with the vertex set $\{v^{j}_{i,1},\hdots, v^{j}_{i,n_j}\},$ $1\leq i\leq n_0(1+n_1)\cdots (1+n_{j-1}).$ Besides, denote the vertices of $G_0$ in $G^{\circledwedge N-1}$ as $\{v^0_{0,1},\cdots, v^{0}_{0,n_0}\},$ and the vertices of $i$-th copy of $G_1$ is given by $\{v^{1}_{i,1},\hdots, v^{1}_{i,n_1}\}$ for $1\leq i\leq n_0.$ Finally, the vertex set of $G^{\circledwedge N-1}$ is given by 
\begin{eqnarray}
   && V^{\circledwedge N-1} \nonumber \\ &=& \left\{v^0_{0,1},\cdots, v^{0}_{0,n_0}\right\} \bigcup \left\{v^1_{i,l} : 1\leq i\leq n_0, 1\leq l\leq  n_1\right\} \bigcup \nonumber \\
    && \left\{v^j_{i,1},\cdots, v^j_{i,n_j} : 2\leq j\leq N-2,\right. \nonumber \\
    && \left. 1\leq i\leq n_0(1+n_1)\cdots (1+n_{j-1})\right\}.\label{eqn:vertexset}
\end{eqnarray}

Now let $d^{(j)}_1,\hdots,d^{(j)}_{n_j}$ denote the degree of the vertices $v^{(j)}_{i,1}, \hdots, v^{(j)}_{i,n_j}$ of the $i$-th copy of the graph $G_j,$ respectively, $0\leq j\leq N-2.$ Then the degree of a vertex $v^{(j)}_{i,s}$ after formation of the nested corona product graph is given by $1+d_s^{(j)}+\sum_{l=j+1}^{N-2} n_l,$ $1\leq s\leq n_j$, where $1$ appears due to the attachment of $G_j^{(i)}$ to an existing vertex of the graph $G^{\circledwedge j}$, and the term $\sum_{l=j+1}^{N-2} n_l$ appears due to the definition of nested corona product graph $G^{\circledwedge N-1}.$

\begin{figure}[htbp]  
\centering  
\subfigure[$G_0=\Pi_0^G=\mathcal{S}^G_2$]  
{  
\begin{tikzpicture}[scale=.8]  
\draw [fill] (0, 0) circle [radius = .1];
\draw [fill] (3, 0) circle [radius = .1]; 
 \draw (0,0) -- (2,0);
 \draw (2,0) -- (3,0);
 \node [below] at (0, 0) {$v^{(0)}_1=I$};
                \node [below] at (3, 0) {$v^{(0)}_2=s_0$};
\end{tikzpicture}  
}  
% The only difference is here, where I have commented out an empty line.
\subfigure[$G_1=\overline{\Pi}_1^G$]  
{  
\begin{tikzpicture}[scale = 0.8]  
 \draw [fill] (0, 0) circle [radius = .1];
\draw [fill] (3, 0) circle [radius = .1];
 \draw (0,0) -- (2,0);
\draw (2,0) -- (3,0);
 \node [below] at (0, 0) {$v^{(1)}_1=s_1$};
\node [below] at (3, 0) {$v^{(1)}_2=s_1s_0$};
\end{tikzpicture}  
}
\subfigure[$G_2=\overline{\Pi}_2^G$]  
{  
\begin{tikzpicture}[scale = 0.8]  
\draw [fill] (0, 0) circle [radius = .1];
\draw [fill] (3, 0) circle [radius = .1];
\draw [fill] (6, 0) circle [radius = .1];
 \draw (0,0) -- (3,0);
\draw (3,0) -- (6,0);
 \node [below] at (0, 0) {$v^{(2)}_1=s_2$};
\node [below] at (3, 0) {$v^{(2)}_2=s_2s_1$};
\node [below] at (6, 0) {$v^{(2)}_3=s_2s_1s_0$};
\end{tikzpicture}  
}
\subfigure[$\mathcal{S}_3^G=\Pi_0^G\circ \overline{\Pi}_1^G$] 
{  
\begin{tikzpicture}[scale = 0.8]  
\draw [fill] (0, 0) circle [radius = .1];
\draw [fill] (3, 0) circle [radius = .1]; 
\draw [fill] (-1, 1) circle [radius = .1];
\draw [fill] (-1, -1) circle [radius = .1];
\draw [fill] (4, 1) circle [radius = .1];
\draw [fill] (4, -1) circle [radius = .1];
 \draw (0,0) -- (3,0);
 \draw[dotted,thick] (0,0) -- (-1,1);
 \draw[dotted,thick] (0,0) -- (-1,-1);
 \draw (-1,-1) -- (-1,1);
  \draw[dotted,thick] (3,0) -- (4,1);
 \draw[dotted,thick] (3,0) -- (4,-1);
 \draw (4,1) -- (4,-1);
 \node [above] at (0, 0) {$I$};
 \node [above] at (3, 0) {$s_0$};
 \node [left] at (-1, 1) {$s_1$};
 \node [left] at (-1, -1) {$s_1s_0$};
 \node [right] at (4, 1) {$s_0s_1$};
 \node [right] at (4, -1) {$s_0s_1s_0$};
\end{tikzpicture}  
}
\subfigure[$\mathcal{S}_4^G=\left(\Pi_0^G\circ \overline{\Pi}_1^G\right)\circ \overline{\Pi}_2^G$] 
{  
\begin{tikzpicture}[scale = 0.8]  
\draw [fill] (0, 0) circle [radius = .1];
\draw [fill] (3, 0) circle [radius = .1]; 
\draw [fill] (-1, 1) circle [radius = .1];
\draw [fill] (-1, -1) circle [radius = .1];
\draw [fill] (4, 1) circle [radius = .1];
\draw [fill] (4, -1) circle [radius = .1];
\draw [fill] (5, 1.5) circle [radius = .1];
\draw [fill] (4.5, 2) circle [radius = .1];
\draw [fill] (4, 2.5) circle [radius = .1];
\draw [fill] (5, -1.5) circle [radius = .1];
\draw [fill] (4.5, -2) circle [radius = .1];
\draw [fill] (4, -2.5) circle [radius = .1];
\draw [fill] (-1, -2.5) circle [radius = .1];
\draw [fill] (-1.5, -2) circle [radius = .1];
\draw [fill] (-2, -1.5) circle [radius = .1];
\draw [fill] (-1, 2.5) circle [radius = .1];
\draw [fill] (-1.5, 2) circle [radius = .1];
\draw [fill] (-2, 1.5) circle [radius = .1];
\draw [fill] (0, 2.5) circle [radius = .1];
\draw [fill] (.5, 2) circle [radius = .1];
\draw [fill] (1, 1.5) circle [radius = .1];
\draw [fill] (3, -2.5) circle [radius = .1];
\draw [fill] (2.5, -2) circle [radius = .1];
\draw [fill] (2, -1.5) circle [radius = .1];
 \draw (0,0) -- (3,0);
 \draw (0,0) -- (-1,1);
 \draw (0,0) -- (-1,-1);
 \draw (-1,-1) -- (-1,1);
  \draw (3,0) -- (4,1);
 \draw (3,0) -- (4,-1);
 \draw (4,1) -- (4,-1);
 \draw (5,1.5) -- (4.5,2);
 \draw (4,2.5) -- (4.5,2);
 \draw[dotted,thick] (4,1) -- (4,2.5);
 \draw[dotted,thick] (4,1) -- (4.5,2);
 \draw[dotted,thick] (4,1) -- (5,1.5);
 \draw (5,-1.5) -- (4.5,-2);
 \draw (4,-2.5) -- (4.5,-2);
 \draw[dotted,thick] (4,-1) -- (4,-2.5);
 \draw[dotted,thick] (4,-1) -- (4.5,-2);
 \draw[dotted,thick] (4,-1) -- (5,-1.5);
  \draw[dotted,thick] (-1,-1) -- (-1,-2.5);
  \draw[dotted,thick] (-1,-1) -- (-1.5,-2);
  \draw[dotted,thick] (-1,-1) -- (-2,-1.5);
  \draw (-1.5,-2) -- (-1,-2.5);
  \draw (-1.5,-2) -- (-2,-1.5);
   \draw[dotted,thick] (-1,1) -- (-1.5,2);
   \draw[dotted,thick] (-1,1) -- (-1,2.5);
   \draw[dotted,thick] (-1,1) -- (-2,1.5);
   \draw (-1,2.5) -- (-1.5,2);
   \draw (-2,1.5) -- (-1.5,2);
   \draw[dotted,thick] (0,0) -- (0,2.5);
   \draw[dotted,thick] (0,0) -- (.5,2);
   \draw[dotted,thick] (0,0) -- (1,1.5);
   \draw (0,2.5) -- (.5,2);
   \draw (.5,2) -- (1,1.5);
   \draw[dotted,thick] (3,0) -- (3,-2.5);
   \draw[dotted,thick] (3,0) -- (2.5,-2);
   \draw[dotted,thick] (3,0) -- (2,-1.5);
   \draw (3,-2.5) -- (2.5,-2);
   \draw (2,-1.5) -- (2.5,-2);
 \node [below] at (0, 0) {$I$};
 \node [above] at (3, 0) {$s_0$};
 \node [left] at (-1, 1) {$s_1$};
 \node [left] at (-1, -1) {$s_1s_0$};
 \node [right] at (4, 1) {$s_0s_1$};
 \node [right] at (4, -1) {$s_0s_1s_0$};
  \node [right] at (0, 2.5) {$s_2$};
  \node [right] at (0.5, 2) {$s_2s_1$};
  \node [right] at (1, 1.5) {$s_2s_1s_0$};
  \node [left] at (2, -1.5) {$s_0s_2s_1s_0$};
  \node [left] at (2.5, -2) {$s_0s_2s_1$};
  \node [left] at (3, -2.5) {$s_0s_2$};
   \node [right] at (5, 1.5) {$s_0s_1s_2$};
   \node [right] at (4.5, 2) {$s_0s_1s_2s_1$};
   \node [right] at (4, 2.5) {$s_0s_1s_2s_1s_0$};
    \node [right] at (5, -1.5) {$s_0s_1s_0s_2$};
   \node [right] at (4.5, -2) {$s_0s_1s_0s_2s_1$};
   \node [right] at (4, -2.5) {$s_0s_1s_0s_2s_1s_0$};
   \node [left] at (-2, -1.5) {$s_1s_0s_2$};
   \node [left] at (-1.5, -2) {$s_1s_0s_2s_1$};
   \node [left] at (-1, -2.5) {$s_1s_0s_2s_1s_0$};
   \node [left] at (-1, 2.5) {$s_1s_2$};
   \node [left] at (-1.5, 2) {$s_1s_2s_1$};
   \node [left] at (-2, 1.5) {$s_1s_2s_1s_0$};
\end{tikzpicture}  
}
\caption{The path graph representation of $\Pi_0^G,$ $\overline{\Pi}_1^G$ and $\overline{\Pi}_2^G$ are depicted in (a), (b) and (c) respectively. The nested corona product graph representation of the symmetric groups $\mathcal{S}_2^G$, $\mathcal{S}_3^G$ and $\mathcal{S}_4^G$ are given by (a), (d) and (e), respectively.}
\label{Fig:gforsym}
\end{figure}
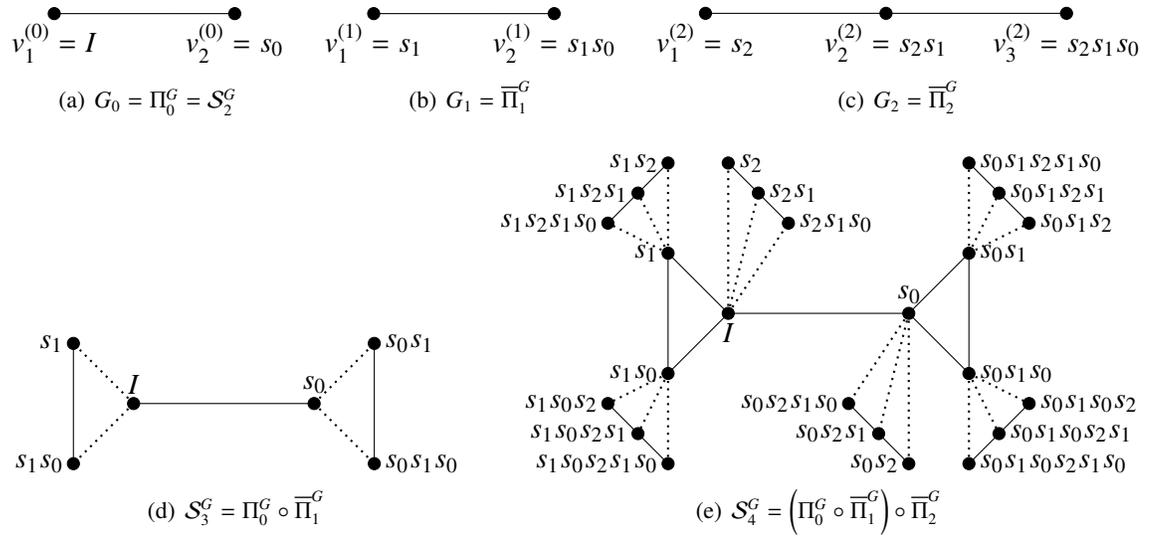

Now, we consider a path graph representation of $\Pi_0=\{I,s_0\}$ and $\overline{\Pi}_k=\Pi_k\setminus \{I\}=\{s_k, s_ks_{k-1}, \hdots, s_ks_{k-1}\cdots s_0\},$ $1\leq k\leq N-2,$ where $\Pi_0^G$ represents path (an edge) with two vertices labelled as $I$ and $s_0,$ whereas $\overline{\Pi}_k^G$ denotes the path graph on $k+1$ vertices labelled as $s_k=v^{(k)}_1,s_ks_{k-1}=v^{(k)}_2,\hdots, s_ks_{k-1}\cdots s_0=v^{(k)}_{k+1}$ with edges between $v^{(k)}_l$ and $v^{(k)}_{l+1},$ $1\leq l\leq k.$ See Figure \ref{Fig:gforsym} for $\overline{\Pi}^G_k,$ $k=1,2,$ and $\Pi^G_0.$ Now we propose a nested corona product graph representation of the symmetric group.  

Indeed, the number of vertices of $ \circledwedge_{j=0}^{N-2} G_j, \,  G_0=\Pi_0, G_j=\overline{\Pi}_j, j\geq 1$ is $N!$ as described by the following corollary. 

\begin{corollary}\label{cor:pg}
 Let $G= \circledwedge_{j=0}^{N-2} G_j, \,  G_0=\Pi_0^G, G_j=\overline{\Pi}_j^G, j\geq 1.$ Then $\left|V(G)\right|=N!.$    
\end{corollary}

\pf Setting $n_0=2$, $n_1=2$ and $n_l=l+1$ for $l\geq 3$ the desired result follows from Theorem \ref{Thm:vencpg}. Indeed \begin{eqnarray*} && n_0(1+n_1)(1+n_2)+\cdots+ (1+n_{N-2}) \\ &=& 2(1+2)(1+3)\cdots (1+N-1) \\ &=& 2\cdot 3\cdot 4\cdots N=N!.\end{eqnarray*} This completes the proof. \hfill{$\square$}

Now we describe the graph $G$ defined in Corollary \ref{cor:pg} as the graph representation of the symmetric group $\mathcal{S}_k$ by assigning the labelling of the vertices as permutations as per the following procedure. 

\begin{algorithm}
  \caption{Nested corona product graph generative model $\mathcal{S}_N^G$ for symmetric group on $N\geq 3$ elements}
  \begin{algorithmic}[1]
    \State The graph representation of $\mathcal{S}_2$ is $\Pi_0^G,$ denoted as $\mathcal{S}^G_2.$
    %\algrule
    \State Label the vertices of $\overline{\Pi}_k^G,$ $k\geq 1$ as the permutations $s_k, s_ks_{k-1}, \hdots, s_ks_{k-1}\cdots s_0.$
    %\algrule[5pt]
    \State When a copy of $\overline{\Pi}_k^G$ is attached to an existing vertex labelled with a permutation $\pi$ in the nested corona product graph then the labelling of the vertices of that $\overline{\Pi}_k^G$ are assigned as $\pi s_k,$ $\pi s_ks_{k-1},$ $\hdots,$ $\pi s_ks_{k-1}\cdots s_0.$
    \State For $N\geq 3,$ $$\mathcal{S}_{N}^G= \circledwedge_{j=0}^{N-2} G_j, \,  G_0=\Pi_0^G, G_j=\overline{\Pi}_j^G, j\geq 1.$$ %represents a network generative model for the symmetric groups. 
  \end{algorithmic}
  \label{Alg:cofcorona}
\end{algorithm}

%\begin{enumerate}
 %   \item The graph representation of $\mathcal{S}_2$ is $\Pi_0^G,$ denoted as $\mathcal{S}^G_2.$
  %  \item Label the vertices of $\overline{\Pi}_k^G,$ $k\geq 3$ as the permutations $s_k, s_ks_{k-1}, \hdots, s_ks_{k-1}\cdots s_0.$
  %  \item When a copy of $\overline{\Pi}_k^G$ is attached to an existing vertex labelled with a permutation $\pi$ in the nested corona product graph $\mathcal{S}^G_{k-2}$ then the labelling of the vertices of that $\overline{\Pi}_k^G$ are assigned as $\pi s_k,$ $\pi s_ks_{k-1},$ $\hdots,$ $\pi s_ks_{k-1}\cdots s_0.$
   % \item For $k\geq 3,$ $$\mathcal{S}_{k}^G= \circledwedge_{j=0}^{k-2} G_j, \,  G_0=\Pi_0^G, G_j=\overline{\Pi}_j^G, j\geq 1,$$ represents a network generative model for the symmetric groups. 
%\end{enumerate}

%Thus the graph representation of $\Pi_0$ and $\Pi_k,$ $k\geq 1$ are given by Figure... with $|V(\Pi_9)|=2$ and $|V(\Pi_k)|=k+1,$ $k\geq 1.$ 

%number of vertices of the path graph of $\Pi_0$ and $\overline{\Pi}_k,$ $k\geq 1$ are $2$ and $k+1$ respectively.  

%First we consider a path graph representation of the non-identity elements of $\Pi_k,$ $1\leq k\leq N-2.$ For $k=2,$ consider $\Pi_0=\{I,s_0\}$ with the path graph on two vertices, labelled by $I$ and $s_0.$ For $k\geq 1,$ the 

In Figure \ref{Fig:gforsym} we provide the nested corona product graph representation of the symmetric groups $\mathcal{S}_k$ for $k=2,3,4.$

\begin{remark}
It should be noted that the choice of representing the set of permutations $\Pi_0$ and $\overline{\Pi}_k,$ $k\geq 1$ by path graphs is ad hoc, and indeed it can be represented by any (simple) graph on $|\Pi_0|$ and $|\overline{\Pi}_k|$ vertices, such as star graph, complete graph or even the null graph with no edges. We justify our choice of the graphs $\Pi_0^G$ and $\overline{\Pi}_k^G,$ $k\geq 1$ as follows:
\begin{enumerate}
    \item If the graph representing $\Pi_0=\{I, s_0\}$ is disconnected then the graph $S^G_k$ would be disconnected for all $k,$ hence the only option to consider an edge graph for $\Pi_0$ so that the resulting graphs remain connected.
    \item The least number of edges in a connected graph on a set of vertices is the path graph, so the choice of path graph for $\overline{\Pi}_k$, $k\geq 1$ does not explode with a large number of edges in $S^G_N.$ Secondly, since the vertices are represented by permutations in $\overline{\Pi}_k,$ the choice of path graph has an algebraic interpretation: any two vertices $\pi$ and $\tau$ are linked by an edge if and only if $\pi=s\tau $ for some adjacent transposition $s.$ 
\end{enumerate}
\end{remark}

Obviously, it follows from the Corollary \ref{cor:pg} and the Algorithm \ref{Alg:cofcorona} that the number of vertices of $\mathcal{S}_N^G$ is $N!,$ which equals the number of permutations on $N$ elements. An interesting property of $S^G_N$ is that it is a union of two isomorphic subgraphs, one originating at $I$ and the other originating at $s_0$, which are joined by the edge $(I,s_0).$  

\subsection{Quantum circuit methods for sampling permutations from $\mathcal{S}_N^G$}

In this section, we show that the circuit model of random sampling of permutations for $n$-qubit systems as discussed in Section \ref{Sec:qcforperm} and the nested corona product graph representation of symmetric groups enable us to sample permutations from specific sets of permutations. Note that permutations on $n$-qubit systems are vertices of the graph $S_{2^n}^G.$ Thus, in this section, we set $N=2^n.$

From equation (\ref{eqn:vertexset}) and Algorithm \ref{Alg:cofcorona} it follows that each vertex of $S_{N}^G$ represents a permutation and any such permutation is decided by a copy of $\overline{\Pi}_j^G,$ $ 1\leq j\leq N-2$ which is attached to an existing vertex. On the other hand, modifying the quantum circuit shown in Figure \ref{fig:sN_rpqc} with each ancillary quantum state $\ket{\Pi_k}_{k+2},$ $k\geq 1$ replaced by $\ket{\overline{\Pi}_k}_{k+1}=\sum_{j=0}^k \frac{1}{\sqrt{k+1}} \ket{j}_{k+1}$ defined as follows, drives a technique for efficient sampling of permutations. 

 \begin{figure}[htbp]
    \centering
     \includegraphics[width=0.50\textwidth]{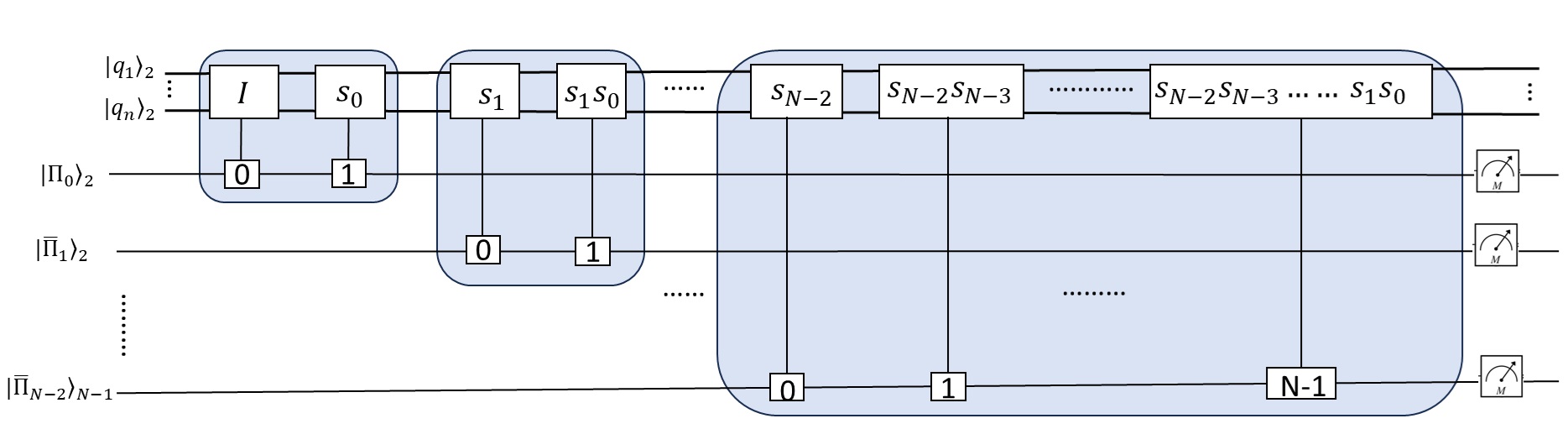} 
\centering \caption {Quantum circuit for sampling of specific permutations}
    \label{fig:circuit_ncorona}
\end{figure}

Below we list how to perform random sampling of permutations through the quantum circuit given by Figure \ref{fig:circuit_ncorona} from some specific subsets of vertices of $\mathcal{S}_N^G.$ 

\begin{enumerate}
    \item Sampling from $\Pi_0$ and $\overline{\Pi}_k,$ $k\geq 1:$ Note that sampling from $\Pi_0$ can be done by measuring $\ket{\Pi_0}_2.$ Note that the permutations in $\overline{\Pi}_k$ for any $k\geq 1$ are represented by the vertices when a copy of $\overline{\Pi}_k^G$ is attached to the vertex $I$ of $\Pi_0^G.$ Thus sampling from nontrivial permutations in $\Pi_k$ can be done by performing a measurement to the state $\ket{\overline{\Pi}_k}.$ 
    
    \item Sampling from $\overline{\Pi}_{l}\overline{\Pi}_m=\{\pi\tau : \pi\in \overline{\Pi}_k, \tau\in \overline{\Pi}_l\}$, $l<m\leq N-2$: The permutations in $\overline{\Pi}_{l}\overline{\Pi}_{m}$  are represented by the vertices which are the vertices of all the copies of $\overline{\Pi}_{m}$ when each of which gets attached to all the vertices of a copy of $\overline{\Pi}^G_l$, which is attached to $I$ (a vertex in $\Pi_0^G$) during the formation of $S^G_k.$ Thus making make a simultaneous measurement of both $\ket{\overline{\Pi}_l}_{l+1}$ and $\ket{\overline{\Pi}_{m}}_{m+1}$ gives the desired result.  
    
   % If $m=l+1$ then note that the permutation in $\overline{\Pi}_{l}\overline{\Pi}_{l+1}$  are given by the vertices which are vertices of all the copies of $\overline{\Pi}_{l+1}$ when they attach to all the vertices of a copy of $\overline{\Pi}^G_l$ which is attached to $I$ (a vertex in $\Pi_0^G$) during the formation of $S^G_k.$ Thus, first perform a measurement of $\ket{\Pi_0}_2,$ upon observing the outcome $\ket{0}_2$ (which corresponds to $I$), make a simultaneous measurement of both $\ket{\overline{\Pi}_l}$ and $\ket{\overline{\Pi}_{l+1}}.$  

   \item Sampling from the set of permutations which are represented by vertices in a copy of $\overline{\Pi}_j^G$ for some $1\leq j\leq N-2$ in $\mathcal{S}_N^G:$  The permutations corresponding to the vertices of a copy of $\overline{\Pi}_j^G$ are of the form $\pi\tau,$ where $\tau\in \overline{\Pi}_j=\{s_j,s_js_{j-1}, \hdots, s_js_{j-1}\cdots s_0\}$ and $\pi$ is the permutation corresponding to the vertex to which $\overline{\Pi}_j^G$ is attached due to the definition of $\mathcal{S}_N^G.$ Now observe that $\pi$ is a vertex of  $\overline{\Pi}_l^G$ for some $l<j,$ and it continues to obtain $\pi=\pi_1\pi_2\cdots \pi_m$ such that $\pi_i\in V(\Pi^G_{k_i})$ for some $k_1<k_2<\hdots <k_m<j$ with $\pi_1\in\{I,s_0\}=V(\Pi_0^G).$ Besides, if the vertex corresponding to $\pi_{k_i}$ in $\overline{\Pi}_{k_i}^G$ represents the $l_i$-th element of $\overline{\Pi}_k^G$ then setting the states $\ket{\overline{\Pi}_{k_i}}_{k_i+1}$ as $\ket{l_i}_{k_i+1}$ in the quantum circuit (Figure \ref{fig:circuit_ncorona}) for $i=1,2,\hdots,m$, the measurement of $\ket{\overline{\Pi}_j}_{j+1}$ will generate a sample from the desired set of permutations. Obviously, any permutation from the desired set will be sampled with probability $\frac{1}{j+1}.$ 

   The equivalent combinatorial interpretation of this method is to identify the shortest path from the vertex $I$ or $s_0$ to the set of vertices chosen to sample the permutations. 

  % \item Sampling from any set $S\subset V(\mathcal{S}_N^G):$
\end{enumerate}

From the above procedures, it is clear that the graph-theoretic interpretation of sampling a set of $k+1,$ $1\leq k\leq N-2$ permutations represented by the set of vertices $\{v^{j}_{i,1}, v^{j}_{i,2}, \hdots, v^{j}_{i,n_j}\},$ as discussed in equation (\ref{eqn:vertexset}) (here $n_j=j+1$ since $G_j=\overline{\Pi}_j^G$) is the identification of the shortest path either from $I$ or $s_0,$ the vertices of $\Pi_0^G$ to the vertex to which the copy of $G_j$ is attached due to the formation of the corona product. Identification of the intermediate vertices of the graphs $\overline{\Pi}_k^G,$ $k<j$ graphs would essentially decide the ancillary states $\ket{\overline{\Pi}_k}_{k+1}$ for executing the circuit simulation for the sampling.

\section{Conclusion}
Exploiting the Steinhaus–Johnson–Trotter algorithm, we present a classical algorithm with time complexity $O(N^2)$ for random sampling of permutations on $N \geq 2$ elements, expressed as products of adjacent transpositions. Building on this, we develop a framework for implementing this algorithm in a quantum circuit model for sampling random permutations from the symmetric group $\mathcal{S}_N$. The proposed quantum circuit model utilizes a primary register of $\lceil \log_2 N \rceil$ qubits and an ancillary register of $(N-1)\lceil \log_2 N \rceil$ qubits. As a result, the asymptotic scaling of the qubit and gate complexities are $O(N \log_2 N)$ and $O(N^3 \log_2 N)$, respectively. This is achieved by constructing quantum circuit representations of adjacent permutations using the $X$ gate, CNOT gate, and $\lceil\log_2N\rceil$-qubit Toffoli gate. Furthermore, we apply this circuit to develop a quantum algorithm for the two-sample randomization test, where $N$ classical data points are encoded in a $\lceil \log_2 N \rceil$-qubit register. The proposed algorithm is shown to reduce the time complexity of the test by a factor of $O(\sqrt{N})$ compared to the classical approach.  
Finally, a nested corona product graph generative model is defined to provide graph representation of symmetric groups, which is also used to define a quantum circuit model for random sampling from specific sets of permutations. \\

\section*{Acknowledgment} The author gratefully acknowledges the anonymous Referees for their insightful suggestions and constructive comments, which have significantly improved both the content and presentation of the paper.

\bibliographystyle{unsrt}
\bibliography{reff}

% biography section
% 
% If you have an EPS/PDF photo (graphicx package needed) extra braces are
% needed around the contents of the optional argument to biography to prevent
% the LaTeX parser from getting confused when it sees the complicated
% \includegraphics command within an optional argument. (You could create
% your own custom macro containing the \includegraphics command to make things
% simpler here.)
%\begin{IEEEbiography}[{\includegraphics[width=1in,height=1.25in,clip,keepaspectratio]{mshell}}]{Michael Shell}
% or if you just want to reserve a space for a photo:

% if you will not have a photo at all:
%\begin{IEEEbiographynophoto}{John Doe}
%Biography text here.
%\end{IEEEbiographynophoto}

% insert where needed to balance the two columns on the last page with
% biographies
%\newpage

%\begin{IEEEbiographynophoto}{Jane Doe}
%Biography text here.
%\end{IEEEbiographynophoto}

% You can push biographies down or up by placing
% a \vfill before or after them. The appropriate
% use of \vfill depends on what kind of text is
% on the last page and whether or not the columns
% are being equalized.

%\vfill

% Can be used to pull up biographies so that the bottom of the last one
% is flush with the other column.
%\enlargethispage{-5in}

% that's all folks
\end{document}